\documentclass[sigconf, nonacm]{acmart}

\usepackage{graphicx}

\usepackage{url}

\usepackage{color}
\usepackage{alltt}
\usepackage{syntax}
\usepackage{newfloat}
\usepackage{mdframed}
\usepackage{multicol}
\usepackage{amsmath}
\usepackage{tikz}
\usepackage{filecontents} 
\usepackage{graphicx}
\usepackage{epstopdf} 
\usepackage{xspace}
\usepackage{enumitem}
\usepackage{svg}
\usepackage{lscape}
\usepackage{longtable}
\usepackage{supertabular}
\usepackage{booktabs}

\usepackage[ruled,linesnumbered,noend]{algorithm2e}
\SetKw{Continue}{continue}
\SetKwInput{KwInput}{Input}
\SetKwInput{KwOutput}{Output}
\SetKwProg{Fn}{Function}{}{end}

\SetCommentSty{mycommfont}

\usepackage{adjustbox}
\usepackage{wrapfig}
\usepackage{graphbox}
\usepackage{rotating}
\usepackage{pdfpages}
\usepackage{mdwlist}
\usepackage{multirow}
\usepackage[T1]{fontenc}

\usepackage{balance}
\usepackage{makecell}

\usepackage[utf8]{inputenc}
\usepackage[UKenglish]{babel}
\usepackage{colorprofiles}
\usepackage{xmpincl}
\usepackage[a-2b,mathxmp]{pdfx}[2018/12/22]
\usepackage{hyperref}

\usepackage[labelfont=bf,skip=5pt,figurename=Fig.]{caption}
\usepackage{subcaption}

\usepackage{hyperref} 

\usepackage[capitalize,nameinlink]{cleveref}
\hypersetup{%
	bookmarksnumbered, bookmarksopen=true, bookmarksopenlevel=1,%
}
\crefname{figure}{Fig.}{Figs.}
\crefname{listing}{Query}{Queries}
\crefname{section}{Section}{Sections}
\crefname{table}{Table}{Tables}
\crefname{BNF}{Grammar}{Grammars}
\crefname{algorithm}{Algorithm}{Algorithms}
\crefname{equation}{Equation}{Equations}

\usepackage{listings}
\definecolor{mygreen}{rgb}{0,0.6,0}
\definecolor{mygray}{rgb}{0.5,0.5,0.5}

\newcommand{\incode}[1]{\lstinline{#1}}

\definecolor{comment-green}{rgb}{0.41,0.6,0.33}
\definecolor{light-gray}{gray}{0.95} 
\lstdefinestyle{myStyleMain}{
    belowcaptionskip=1\baselineskip,
    breaklines=true,
    keywordstyle=\color{purple},
    commentstyle=\color{comment-green},
    morekeywords={where, exclude, search, backtrack, forwardtrack, include, limit, display, like},
    backgroundcolor=\color{light-gray},  
    xleftmargin=8pt,
}

\DeclareFloatingEnvironment[
fileext   = logr,
listname  = {List of Grammars},
name      = Grammar,
placement = !htbp
]{BNF}

\newcommand{\msim}{\raise.17ex\hbox{$\scriptstyle\sim$}}

\newcommand{\myparatight}[1]{\smallskip\noindent{\bf {#1}.}}

\newcommand{\eat}[1]{}
\newcommand{\eg}{e.g.,\xspace}
\newcommand{\ie}{i.e.,\xspace}

\usepackage{pifont}
\newcommand{\circled}[1]{\ding{\numexpr201+#1\relax}}
\newif\ifshowcomment
\showcommenttrue

\ifshowcomment
    \newcommand{\pgao}[1]{\textsf{\color{red}{({Peng: #1})}}}
    \newcommand{\saimon}[1]{\textsf{\color{orange}{({Saimon: #1})}}}
    \newcommand{\rev}[1]{\textsf{\color{purple}{({Comment: #1})}}}
    \newcommand{\revdone}[1]{\textsf{\color{blue}{({Rev-done: #1})}}}
\else
    \newcommand{\pgao}[1]{}
    \newcommand{\saimon}[1]{}
    \newcommand{\rev}[1]{}
    \newcommand{\revdone}[1]{}
\fi

\newcommand{\tool}{\textsc{Provexa}\xspace}
\newcommand{\toolno}{\textsc{Provexa}}
\newcommand{\lang}{\textsc{ProvQL}\xspace}
\newcommand{\langno}{\textsc{ProvQL}}

\usepackage{soul}

\begin{document}

\date{}

\title{Enabling Efficient Attack Investigation via Human-in-the-Loop Security Analysis}

\author{Saimon Amanuel Tsegai}
\affiliation{%
  \institution{Virginia Tech}
    \country{}
}
\email{saimon.tsegai@vt.edu}

\author{Xinyu Yang}
\affiliation{%
  \institution{Virginia Tech}
  \country{}
}
\email{xinyuyang@vt.edu}

\author{Haoyuan Liu}
\affiliation{%
  \institution{University of California, Berkeley}
    \country{}
}
\email{hy.liu@berkeley.edu}

\author{Peng Gao}
\affiliation{%
  \institution{Virginia Tech}
    \country{}
}
\email{penggao@vt.edu}

\begin{abstract}

System auditing is a vital technique for collecting system call events as system provenance and investigating complex multi-step attacks such as Advanced Persistent Threats. However, existing attack investigation methods struggle to uncover long attack sequences due to the massive volume of system provenance data and their inability to focus on attack-relevant parts. 
In this paper, we present \tool, a defense system that 
enables human analysts to effectively analyze large-scale system provenance to reveal multi-step attack sequences. \tool introduces an expressive domain-specific language, \lang, that offers essential primitives for various types of attack analyses (\eg attack pattern search, attack dependency tracking) with user-defined constraints, enabling analysts to focus on attack-relevant parts and iteratively sift through the large provenance data. Moreover, \tool provides an optimized execution engine for efficient language execution. Our extensive evaluations on a wide range of attack scenarios demonstrate the practical effectiveness of \tool in facilitating timely attack investigation.

The source code, data, and other artifacts have been made available at \url{https://github.com/peng-gao-lab/Provexa}.

\end{abstract}

\maketitle

\section{Introduction}
\label{sec:intro}

Despite significant increases in spending on operations security, the frequency of modern targeted cyberattacks, such as Advanced Persistent Threats (APTs), continues to rise.
Unlike traditional threats, these attacks are highly sophisticated, leveraging multiple vulnerabilities to infiltrate the system and exfiltrate sensitive data through a series of steps~\cite{milajerdi2019holmes}. Consequently, many high-profile businesses have suffered massive data breaches and losses~\cite{breach21,target,equifax}.

To counter these \emph{intrusive multi-step attacks}, ubiquitous system auditing has emerged as a vital approach for monitoring attack footprints~\cite{auditd}.
System auditing monitors system call events between system entities as system audit logs. Unlike application-level monitoring (\eg Apache server logging), 
which only provides limited knowledge about specific applications and generates logs in different formats, system auditing is not tied to applications and generates audit logs with a consistent structure. The collected audit logs further enable the construction of a system provenance graph~\cite{inam2023history}, in which nodes represent system entities (\eg processes, files, network sockets) and edges represent their system call events (\eg a process writes to a file). A system provenance graph provides a holistic view of all the activities in the system, which is particularly useful for investigating cyber attacks and uncovering attack steps~\cite{backtracking,liu2018priotracker}.
However, system auditing produces a huge amount of daily logs (\eg 0.5 GB $\sim$ 1 GB for one enterprise host~\cite{reduction}), resulting in a giant provenance graph. Additionally, the complexity of multi-step attacks poses significant challenges to existing system provenance-based defenses, which struggle to effectively uncover long attack sequences within such large provenance graphs.

\begin{figure*}[t]
    \centering
    \includegraphics[width=.95\linewidth]{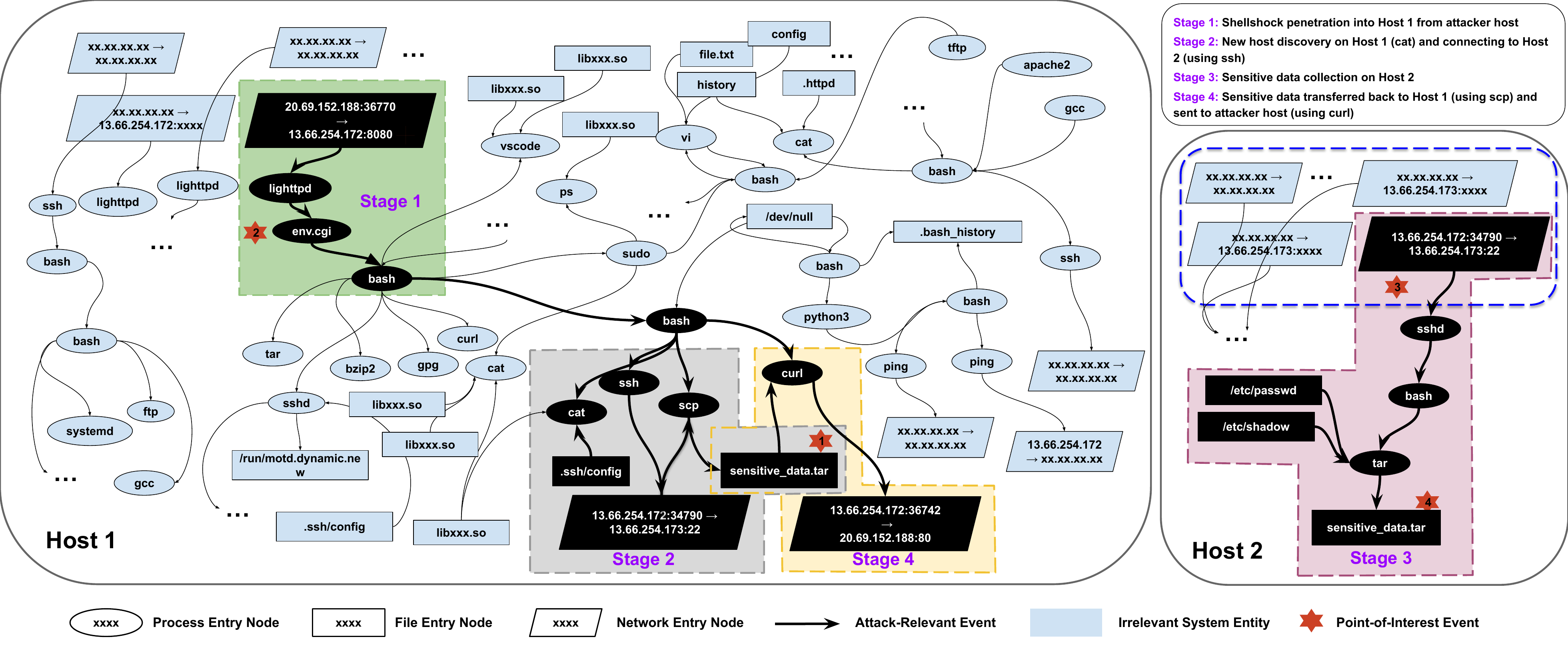}
    \caption{System dependency graphs for a multi-stage, multi-host data leakage attack. 
    The combined dependency graphs of the two victim hosts contain $100,524$ nodes and $154,353$ edges. The attack-relevant nodes and edges, highlighted in dark black, comprise only 20 nodes and 20 edges, indicating the significant challenge of finding ``a needle in a haystack''.
    }
    \label{fig:demo}
\end{figure*}

Existing provenance-based attack investigation approaches often leverage causal dependency tracking~\cite{backtracking,backtracking2, liu2018priotracker,fang2022backpropagating,kwon2018mci,ma2016protracer,lee2013high}.
These approaches model the control/data flow dependencies between system entities in a system event, track the dependencies from a Point-of-Interest (POI) event (\eg an alert event like a process creating a suspicious file), and construct a dependency graph, which is a subgraph of the whole system provenance graph. Security analysts can inspect the dependency graph to reveal the attack sequence by reconstructing the chain of events that lead to the POI event. 
However, due to \textbf{\emph{the lack of fine-grained user control}} of the tracking process, these approaches suffer from dependency explosion: the generated dependency graph is gigantic (containing >100K edges) and contains many system events that are \emph{irrelevant} to the attack (\eg events that load irrelevant system libraries). The problem is worse for multi-step attacks with long attack sequences, making it hard for security analysts to sift through the graph and identify the attack-relevant parts. \cref{fig:demo} shows the dependency graph of a multi-stage data leakage attack and illustrates this problem: a small number of attack-relevant events are buried in an overwhelmingly large number of irrelevant events.

\myparatight{Goal and challenges}
We aim to design and build a new defense system that (1) effectively filters out irrelevant system events and reveals complex attack sequences, and (2) efficiently analyzes large-scale system provenance data for a timely investigation. 
As reported by many leading security vendors~\cite{ibm-threat-report,crowdstrike-threat-report}, \textbf{\emph{human-in-the-loop security}} emerges as a key paradigm for attack investigation. Humans bring unique strengths to cybersecurity, such as the ability to interpret subtle patterns, understand context, and make nuanced decisions that are difficult for automated systems alone.
Such domain knowledge on expected system behaviors and malicious event patterns is crucial for filtering out irrelevant events and reducing dependency explosion as attacks become more sophisticated. 
Besides, attack investigation is an \textbf{\emph{iterative process}} that involves multiple rounds of data exploration. 
An effective defense should provide a flexible and expressive interface for human analysts to incorporate the knowledge and customize the defenses for various attacks.

While several defenses have attempted to identify attack-relevant events from system provenance graphs, they have notable limitations.
Multiple approaches seek to enhance dependency tracking fidelity through algorithms based on heuristic rules~\cite{liu2018priotracker,fang2022backpropagating,hassan2019nodoze} that lead to information loss, or through binary instrumentation and kernel customization~\cite{ma2016protracer,kwon2018mci} that introduce intrusive system changes.
Several other works aim to detect malicious activities through subgraph matching, employing non-learning-based~\cite{milajerdi2019poirot} and learning-based~\cite{altinisik2023provg,wei2021deephunter} techniques. 
However, these approaches are computationally expensive, requiring extensive offline model training or incurring high runtime overhead over large provenance graphs. Learning-based methods also face generalization challenges, as their models can quickly become outdated with evolving system behaviors.
Moreover, existing approaches overlook the importance of human-in-the-loop investigation, lacking flexible and efficient mechanisms to incorporate expert knowledge iteratively. Most methods assume investigations can be completed in a single round, \textbf{\emph{providing limited support for step-by-step inquiries}}—an essential process for thoroughly investigating complex attacks.

\myparatight{Contributions}
In this work, we take an orthogonal approach to existing solutions. We introduce \tool ($\sim$15K LOC), a system that facilitates practical investigation of complex attacks through efficient human-in-the-loop provenance analysis.
\tool leverages system auditing frameworks and databases for provenance collection and storage. Its core contribution lies in the design of a powerful domain-specific language (DSL), called  Provenance Query Language (\lang), which enables iterative, human-in-the-loop investigations over large-scale system provenance data.
\lang treats system entities and events as first-class citizens and offers two primary query syntaxes for key investigative tasks: (1) \emph{attack pattern search:} searching for complex event patterns indicative of malicious behaviors; (2) \emph{attack dependency tracking:} tracking causal dependencies between events to uncover attack sequences and entry points.
\lang provides rich constructs that allow users to constrain the search and tracking space to focus on critical parts while filtering out noise, mitigating dependency explosion.

\lang is expressive and intuitive to use. 
With its high-level, declarative syntax, \lang abstracts away the low-level complexities of data storage across different backends, enabling security analysts to focus on core attack behaviors rather than low-level details such as table joins. 
The two syntaxes offer complementary attack investigation capabilities. Search queries help locate potential suspects (\eg POI events) for tracking and identify malicious events without dependencies. Tracking queries can uncover long dependency chains that search queries cannot express.
To further facilitate iterative investigations,  \lang allows intermediate query results to be bound to variables and reused in subsequent queries, enabling analysts to refine their findings progressively.

Attack investigation is a time-critical mission to prevent further damage~\cite{liu2018priotracker}.
To efficiently execute \lang queries over large system provenance, \tool employs a domain-aware query scheduler that decomposes the \lang query into small data queries and schedules their execution based on their pruning power, semantic dependencies, and domain characteristics. 
\tool also features an in-memory management technique that maintains intermediate results bound to variables as in-memory graphs for subsequent manipulations and querying. Queries executing in this mode are much faster, facilitating iterative investigations.

\myparatight{Evaluation}
We deployed \tool on a testbed and extensively evaluated its effectiveness in reducing dependency explosion, uncovering attack sequences, and executing \lang queries efficiently. 
Additionally, we assessed \langno’s usability and effectiveness through a user study.
To conduct a thorough evaluation, we built a comprehensive benchmark by executing a wide range of attack scenarios on our testbed and collecting millions of real system events.
We compared \tool with multiple baselines, including state-of-the-art provenance-based attack investigation methods (BackTracker~\cite{backtracking}, PrioTracker~\cite{liu2018priotracker}, DepImpact~\cite{fang2022backpropagating}), general-purpose query languages (SQL~\cite{sql}, Cypher~\cite{cypher}), and a widely used industry attack investigation solution (Splunk~\cite{splunk}).

The results show that:
(1) \tool can accurately uncover the attack sequence in all scenarios while reducing dependency explosion, achieving 0.8766 F1-score and 58,991$\times$ graph reduction rate. 
This significantly outperforms existing defenses; BackTracker, PrioTracker, and DepImpact have only 0.2526, 0.2526, and 0.2604 F1-score, and 9$\times$, 42$\times$, and 24$\times$ reduction rate, respectively. 
(2) \tool can efficiently execute \lang queries over massive provenance, running much faster than SQL and Cypher queries.
(3) \lang features shorter query length with fewer constructs compared to SQL, Cypher, and Splunk, making investigation logic easier to express and maintain.
(4) Our user study confirms \langno's advantages to streamline investigation workflows, reduce cognitive load, and enhance user experience.
These results demonstrate that \toolno's superiority in combating sophisticated attacks.

\section{Background}
\label{sec:background}

\myparatight{Causal dependency tracking}
Causal dependency tracking infers dependencies of system events and presents the dependencies as a directed dependency graph.
In the dependency graph $G(V, E)$, a node $v\in V$ denotes a system entity (\eg process, file, or network socket).
An edge $e(u,v)\in E$ denotes a system event that involves two entities $u$ and $v$ (\eg process creation, file read or write, and network access).
Edge direction (from source node $u$ to sink node $v$) indicates the information flow direction.
For example, for a process reading data from a file, the file is the source and the process is the sink.
For a process writing data to a file, the process is the source and the file is the sink.
Each edge $e(u, v)$ is associated with a time window, $[ts(e), te(e)]$, where $ts(e)$ and $te(e)$ denote the start/end time of the event $e$.
We adopt the definition consistent with previous studies~\cite{backtracking,backtracking2, liu2018priotracker,fang2022backpropagating,kwon2018mci,ma2016protracer,lee2013high} to infer edge directions for different systems calls and event causal dependencies.
Formally, for two events $e_1(u_1, v_1)$ and $e_2(u_2, v_2)$ (suppose $e_1$ occurs earlier than $e_2$), they have causal dependency if $v_1 = u_2$ and $ts(e_1) < te(e_2)$.

Causal dependency tracking, introduced in the BackTracker~\cite{backtracking}, enables two important security analyses: (1) \emph{backward tracking} that identifies attack entry points, and (2) \emph{forward tracking} that investigates attack ramifications.
Given a POI event $e_{poi}(u,v)$, a backward tracking traces back from the source node $u$ to find all events that have causal dependency on $u$,
and a forward tracking traces forward from the sink node $v$ to find all events on which $v$ has causal dependency.
The output of backward/forward tracking is a backward/forward dependency graph. 
Take backward tracking as an example. Starting from an empty queue $Q$ and an empty graph $G$, we first add the POI event to $Q$ and $G$. Then, we iteratively remove an event $e$ from $Q$, find other events that have dependencies on $e$, and add these events to $Q$ and $G$. We repeat this process until $Q$ becomes empty, and the final $G$ is the backward dependency graph. 
A major challenge in dependency tracking is dependency explosion. 
As shown in \cref{subsubsec:eval-rq1}, \emph{the dependency graphs produced remain excessively large and often miss critical attack-relevant events.}

\myparatight{Challenge: Needle in a haystack}
\cref{fig:demo} illustrates an example multi-stage data leakage attack across three hosts (one attacker host, two victim hosts Host 1 and 2)~\cite{cve}.
The attacker (\incode{20.69.152.188}) identifies a victim Host 1 (\incode{13.66.254.172}) in an enterprise network that is vulnerable to the \incode{lighttpd} Shellshock~\cite{shellshock} exploit.
To steal data from the victim Host 1, the attacker launches a series of attack steps divided into four stages:
(1) \emph{Stage 1:} The attacker leverages Shellshock vulnerability in the \incode{lighttpd} web server on Host 1 to penetrate into Host 1 from the attacker host.
(2) \emph{Stage 2:} The attacker discovers all the reachable hosts of Host 1 and connects to the discovered one, Host 2, using \incode{ssh}.
(3) \emph{Stage 3:} The attacker uses \incode{tar} to pack the sensitive files (\eg \incode{/etc/passwd}, \incode{/etc/shadow}) on Host 2 as \incode{sensitive\_data.tar}.
(4) \emph{Stage 4:} The attacker uses \incode{scp} on Host 1 to fetch the data \incode{sensitive\_data.tar} from Host 2. The attacker then uses \incode{curl} on Host 1 to send the sensitive data back to the attacker host.
We can observe that attack investigation faces a significant challenge of finding ``a needle in a haystack'':
a small number of critical attack-relevant edges ($20$; colored in dark black) are buried in an overwhelmingly large number ($154$K) of irrelevant edges.
The same imbalance observation holds for attack-relevant nodes ($20$ vs. $101$K). 
This makes attack investigation particularly challenging.

\section{System Overview}
\label{sec:overview}

\begin{figure}[t]
    \centering
    \includegraphics[width=\linewidth]{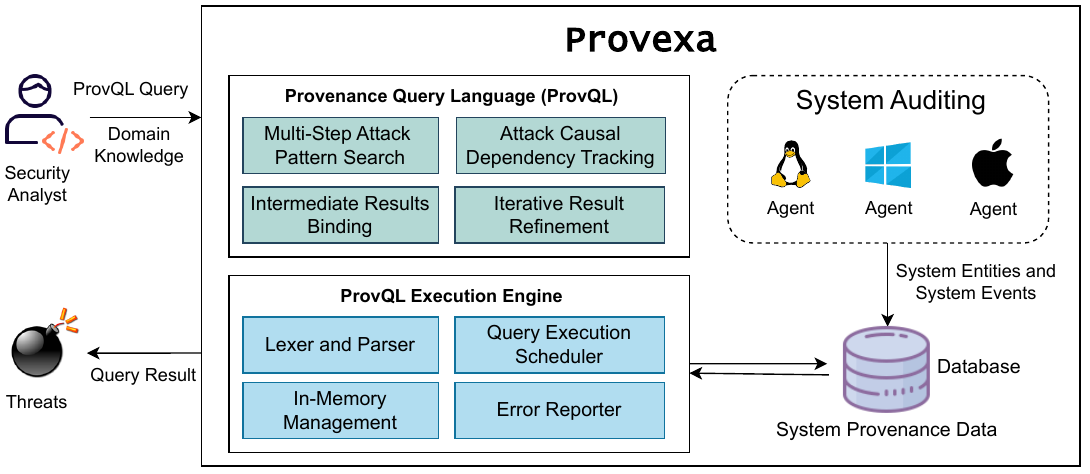}
    \caption{Architecture of \tool}
    
    \label{fig:arch}
    \vspace{-1ex}
\end{figure}

\cref{fig:arch} illustrates the system architecture. \tool uses monitoring agents deployed on hosts to collect audit logs. \tool then parses the logs into a sequence of system events among system entities and sends the parsed data to the database for storage. 
On top of the database, \tool provides a DSL, \lang, for investigating attack behaviors.
\lang integrates a collection of constructs for attack pattern search and attack dependency tracking analyses, as well as constructs for iteratively refining intermediate investigation results.
Given a \lang query input by the security analyst, the language parser performs syntax analysis and semantic analysis of it. The execution scheduler then generates an execution plan based on our specialized optimizations and schedules the execution. 

To enhance the user experience in iteratively exploring system provenance, we developed a user interface (UI) on top of our query execution engine.
Inspired by computational notebooks like Jupyter Notebook, our UI features code cells and inline outputs.
Users can construct queries within these cells. Executing a cell displays the corresponding provenance graphs, and users can build upon results from previous cells to progressively deepen their investigation. Executed cells also serve as a record of the investigation process.
A demo video showing how users can leverage our UI to iteratively investigate complex attacks is available at our project website~\cite{provexa-website}.

Our threat model is similar to that of many previous works on system auditing~\cite{backtracking,backtracking2, liu2018priotracker,fang2022backpropagating,kwon2018mci,ma2016protracer,lee2013high}.
We assume the presence of an attacker seeking to attack the system from outside: the attacker may seek to access or modify unauthorized resources, exfiltrate confidential data, or install and spread malware.
Our trusted computing base includes OS kernels, host agents, data storage, and query execution engine. 
We assume that OS kernels and collected audit logs are secure from compromise. 
We do not consider malicious administrators who can disable the host agent or tamper with system audit logs, or implicit information flows like covert and timing channels which do not go through kernel-layer auditing.

\section{System Auditing Infrastructure}
\label{sec:infrastructure}

\myparatight{Data model and monitoring agents}
System auditing collects system-level events about system calls from the OS kernel. These events describe the interactions among various system entities.
As shown in previous studies~\cite{hassan2019nodoze,ji2017rain,backtracking,liu2018priotracker,ma2016protracer}, on mainstream OSes, system entities are primarily \emph{files, processes, and network sockets}, and system calls are mapped to three major types of system events: (1) file access, (2) processes creation and destruction, and (3) network access. 
Thus, in our data model, we primarily consider these system entities. 
We consider a system event as the interaction between two system entities represented as $\langle$subject\_entity, operation\_type, object\_entity$\rangle$.
Subject entities are processes from software applications and objects can be files, processes, and network sockets.
We categorize system events into three types according to the types of their object entities: file events, process events, and network events.

We develop monitoring agents using system auditing frameworks for different OSes: Sysdig~\cite{sysdig} for Linux,  Procmon~\cite{procmon} for Windows. Deployed on each host, our agent continuously monitors system activities, collects system audit logs, and extracts attributes critical for security analysis.
\cref{tab:entity-attributes,tab:event-attributes} show representative entity and event attributes that our agent extracts.
Following previous works~\cite{liu2018priotracker,hassan2019nodoze,fang2022backpropagating}, to uniquely identify entities, for a process entity, we use the process executable name and PID as its identifier. For a file entity, we use the absolute path as its identifier. For a network socket entity, we use the 5-tuple (source/destination IP, source/destination port, protocol) as its identifier.

\begin{table}[t]
\centering
\caption{Representative entity attributes collected}
\begin{adjustbox}{width=\linewidth}
    \begin{tabular}{ll}
      	\toprule
			\textbf{System Entity}     & \textbf{Attributes} \\ \midrule
			File                & Name, Path, User, Group \\
			Process             & PID, Executable Name, User, Group, Command Line Args \\ 
			Network Socket      & SRC/DST IP, SRC/DST Port, Protocol \\ \bottomrule
    \end{tabular}
\end{adjustbox}
\label{tab:entity-attributes}
\end{table}

\begin{table}[!t]
	\centering
 	\caption{Representative event attributes collected}
	\begin{adjustbox}{width=\linewidth}
		\begin{tabular}{l|l}
			\toprule
            \textbf{Type}      & Operation Type (\eg Read/Write/Execute/Start/End/Rename) \\
			\textbf{Time}		    & Start Time, End Time, Duration \\
			\textbf{Misc.}		    & Subject\_ID, Object\_ID, Data Amount \\ \bottomrule
		\end{tabular}
	\end{adjustbox}
	\label{tab:event-attributes}
\end{table}


\myparatight{Data storage}
\tool stores the parsed system entities and system events in the databases so that the collected provenance data can be persisted.
The current implementation of \tool supports two types of data storage backends: relational database PostgreSQL~\cite{postgresql} and graph database Neo4j~\cite{neo4j}.
This enables \tool to leverage the services these mature infrastructures provide, such as data management, indexing mechanism, querying, and data recovery.
We perform data normalization to reduce redundancy and improve maintainability.
In PostgreSQL, \tool stores system entities and events across six separate tables, with attributes organized into columns (i.e., file, process, and network entity tables; file event, process event, and network event tables).
In Neo4j, \tool represents system entities as nodes and system events as edges (i.e., file, process, and network entity nodes; file event, process event, and network event edges).
Entities can be linked to events by matching the Entity\_ID attribute with the Subject/Object\_ID attributes of events. Indexes are created on key attributes (\eg file name, process executable name, source/destination IP) to speed up the search.

\section{\lang Language Design}
\label{sec:language}

\lang integrates critical primitives for attack investigation, supporting two major types of analysis: multi-step attack pattern search and attack dependency tracking. \cref{bnf:grammar} presents the grammar.

\subsection{Multi-Step Attack Pattern Search}
\label{subsec:search-syntax}
The search syntax (\ie Rule $\langle search\_stmt\rangle$ in \cref{bnf:grammar}) allows users to specify multiple events with constraints on entity/event attributes or event relationships.
This enables the search for multi-event patterns that represent complex, multi-step attack behaviors.

Take Query \circled{1} in \cref{subsec:query-cases} as an example.
First, we specify a database (\eg \incode{db(host1)}) as the data source of the search.
Next, we define three system entities with constraints on their types and attributes (\eg \incode{e1\{name="curl", type=process\}}). Then, we define two system events using the three entities (\eg \incode{e2[read]->e1}, where the  arrow indicates information flow direction) based on Rule $\langle search\_rel\_expr \rangle$.
Besides the \emph{structural relationship} that two events are connected by the same entity \incode{e1}, we constrain their \emph{temporal relationship}: the two events occur within one second (\eg \incode{&&[<1s]}).
Together, these three entities and two events define a subgraph pattern depicting the attack behavior: using \incode{curl} to transfer a sensitive \incode{tar} file to an IP. The transfer is carried out by first reading data from the file and then writing the data to the socket.
After defining the multi-event pattern, we bind the results to a variable (\eg \incode{poi1}), indicating the results are retained in memory.
This variable can be used in subsequent queries (\eg results of Query \circled{1} are used as start entities for tracking in Query \circled{2}) for further refinement.
Users can also visualize the results in our UI by displaying this variable.

\subsection{Attack Dependency Tracking}
\label{subsec:tracking-syntax}

The tracking syntax (\ie Rule $\langle track\_stmt\rangle$) provides constructs for various types of fine-grained control of the causal dependency tracking process, including: tracking direction (backward/forward), POI, optional tracking depth, and entity and event constraints. 
This enables security analysts to effectively mitigate the dependency explosion during tracking by prioritizing attack-relevant parts.

Take Query \circled{2} in \cref{subsec:query-cases} as an example. 
A backward tracking query is specified with a database (\eg \incode{db(host1)}) as the data source.
The tracking begins with the execution results of the previous query, where \incode{poi1} is the starting entity, which was defined in Query \circled{1} and bound to its results.
To constrain the process, Query \circled{2} excludes entities with the name \incode{vscode} (\ie Rule $\langle track\_filter\rangle$), which are most likely benign. 
The execution results of Query \circled{2} produce a backward dependency graph, which is stored in the variable \incode{g1} for display in the UI and can be used in subsequent queries.

\begin{center}
\begin{minipage}{\linewidth}
\footnotesize
\begin{mdframed}
    \setlength{\grammarparsep}{0pt} 
    \setlength{\grammarindent}{8em} 
    \lit*{}  
    \begin{grammar}
    
    <start> ::= (<search_stmt> | <track_stmt> | <graph_op_stmt> | <display_stmt>)*
    
    \end{grammar}
    
    \vspace{1pt}

    \begin{grammar}
    
    <search_stmt> ::= (<id> `=')? `search' `from' <data_source> `where' <search_node> `with'  <search_rel_expr> `;' ;  
    
    <search_node> ::= <id> `{' <expr> `}'
    \alt <search_node> `,' <search_node>
    
    <data_source> ::= <id>

    <search_rel_expr> ::= <search_rel>
    \alt <search_rel> <search_rel_op> <search_rel>

    <search_rel> ::= <id> (`[' event_op `]')? `->' <id>

    <search_rel_op> ::= `&&' (`[' <int> (`m' | `s' | `ms') `]')?
    \alt `||';

    <expr> ::= `(' <expr> `)'
    \alt `!' <expr>
    \alt <expr> (`&&' | `||') <expr>
    \alt <binary_expr>
    
    <binary_expr> ::= <string_attribute> <eq_op> (<string> | <entity_type_id>)
    \alt <numerical_attribute> <numerical_op> <int>
    
    <string_attribute> ::= `type' | `name' | `path' | `dstip' | `srcip' | `exename' | `exepath' | `cmdline' | `optype'
    
    <numerical_attribute> ::= `id' | `srcid' | `dstid' | `starttime' | `endtime' | `amount' | `pid' | `srcport' | `dstport'
    
    <event_op> ::= `read' | `write' | `execve' | `readv' | `writev' | `rename' | `fork' | `clone' | `recvfrom' | `sendto'
    
    <entity_type_id> ::= `process' | `file' | `network'
    
    <eq_op> ::= `=' | `!=' | `like'
    
    <numerical_op> ::= <eq_op> | `>' | `>=' | `\textless' | `\textless='

    \end{grammar}
    
    \vspace{1pt}

    \begin{grammar}
    <track_stmt> ::= (<id> `=')? <track_direction> `track' <track_poi> `from' <data_source> <track_filter> <track_limit>? `;' ;

    <track_direction> ::= `back' | `forward'

    <track_poi> ::= (<track_constraint_expr> | <id>)
    
    <track_filter> ::= (`include' <type_track_filter>)? (`exclude' <type_track_filter>)?

    <type_track_filter> ::= (`nodes' <track_constraint_expr>)? (`,')? (`edges' <track_constraint_expr>)?
    
    <track_constraint_expr> ::= `where' <expr>

    <track_limit> ::= `limit' (`step' <int>)? (`,')? (`time' <int> `s|min')?
    \end{grammar}
    
    \vspace{1pt}

    \begin{grammar}
    
    <graph_op_stmt> ::= <id> `=' <graph_expr> `;'
    
    <display_stmt> ::= `display' <graph_expr> `;'

    <graph_expr> ::=  <id>
    \alt `(' <graph_expr> `)'
    \alt <graph_expr> <graph_op> <graph_expr>

    <graph_op> ::= `|' | `\&' | `-'
    \end{grammar}
\end{mdframed}
\captionof{figure}{BNF grammar of \lang}
\label{bnf:grammar}
\end{minipage}
\end{center}


\subsection{Intermediate Results Binding}
\label{subsec:binding}

\lang queries return a \emph{subgraph} of system events as results.
Search queries return a subgraph that matches the specified multi-event pattern, while tracking queries return a subgraph aligned with the dependency tracking constraints.
Designed to help security analysts iteratively refine their investigation results, \lang supports \emph{binding query results to variables} and retaining them in memory. This enables analysts to efficiently access and modify intermediate results through subsequent queries without reprocessing the entire dataset. This approach significantly reduces query execution time.
For example, Query \circled{4} in \cref{subsec:query-cases} defines the variable \incode{g3} and binds it to the tracked subgraph. Query \circled{5} then uses \incode{g3} as the data source for its search, i.e., search operates on the in-memory \incode{g3} rather than retrieving data from the database again.

To flexibly manipulate results, \lang supports different in-memory graph operations through variables. In Query \circled{6}, a union operation (\incode{g1 | g3}) merges the previously identified segments of the attack sequence, with the resulting subgraph bound to \incode{g4}. Operations such as graph intersection and difference are also supported.

\section{\lang Execution Engine}
\label{sec:execution}

A straightforward but inefficient approach to executing a \lang query is to translate the \emph{entire} query into a single semantically equivalent SQL or Cypher query for execution.
Search queries can be translated into standard SQL or Cypher queries involving multiple joins and constraints.
For example, a single-step \lang query that finds a \incode{curl} process reading from a \incode{tar} file can be translated into a \incode{SELECT} query joining the process, file, and process event tables, with filters on process executable and file names.
Tracking queries are inherently recursive, as they traverse causality chains to find all related past or future events. These can be translated into recursive SQL queries using the \incode{WITH RECURSIVE} clause, or Cypher queries that match variable-length paths with filters.

However, this approach becomes inefficient when querying complex attack patterns over large system provenance. General query optimizers are not tailored to the system provenance domain.
Multi-step search queries will translate into large SQL or Cypher queries with numerous joins and constraints, which could lead to unpredictable performance.
Tracking queries face similar issues: recursive SQL requires full table scans and joins, while Cypher exhaustively matches all paths before filtering, as it lacks on-the-fly constrained tracking~\cite{neo4j-cannot-track-with-filter}.
As analyzed in \cref{subsubsec:eval-rq3}, this single translation strategy often results in inefficient query plans.

\myparatight{Domain-aware query scheduler}
Rather than compiling an entire \lang query into a single inefficient SQL or Cypher query and relying on generic, domain-agnostic schedulers provided by underlying databases, we design a custom scheduler tailored to our domain.
Our scheduler decomposes the query into smaller retrieval steps, each translated into a small SQL/Cypher query. 
This decomposition leverages domain-specific observations, such as common patterns in system provenance data, to identify meaningful subqueries that are efficient to execute.
The resulting small queries are then executed in an order informed by the semantics and structure of the provenance domain, taking into account estimated pruning power and query dependencies to maximize overall efficiency.

For a multi-step search query, each event pattern (e.g., \incode{e2[read]->e1}, \incode{e1[write]->e3} in Query~\circled{1}, indicating a single step) is translated into a small SQL or Cypher query. The SQL version joins two entity tables and one event table with constraints in the WHERE clause; the Cypher version matches two entity nodes and one event edge with filters.
To prioritize execution, the scheduler assigns each pattern a pruning score based on its semantics--a higher score indicates greater potential to reduce the search space. 
In our implementation, we use a coarse heuristic: patterns with more constraints receive higher scores. 
Additionally, inspired by prior observations that file events are more frequent than process or network events~\cite{reduction}, we assign lower scores to file event patterns.
This approach is quick to compute and proves effective in practice, as shown in ~\cref{subsubsec:eval-rq2}.

When scheduling execution, the scheduler prioritizes event patterns with higher pruning scores and enforces an order based on dependencies. For instance, if two event patterns are connected through the same entity, the scheduler executes the higher-scoring data query first and incorporates the returned event IDs as additional WHERE clause constraints in the other data query. This significantly reduces the search space of the second query.

For a tracking query, the scheduler maintains a queue during recursive causality tracking.
For each event dequeued, it compiles its causal dependency conditions--such as information flow, timestamps, and constraints--into small, non-recursive SQL or Cypher queries with appropriate WHERE filters.
These small queries are executed to retrieve all dependent events and recursively update the queue. 
A key advantage of this approach is that it enables efficient \emph{on-the-fly} constrained tracking, avoiding the overhead of retrieving all dependencies upfront and filtering afterward.

Note that \toolno's scheduler does not replace existing database schedulers and optimizations but acts as an additional optimization layer for the system provenance domain. Each intermediate data query is executed on the databases. 
Due to their small sizes, their execution is typically fast, and can benefit from the indexing and database optimizations. 
Importantly, our query decomposition strategy leverages domain knowledge of system provenance to guide the breakdown of queries into semantically meaningful investigative steps rather than applying arbitrary splitting. This design aligns with how real-world investigations unfold: each step typically examines a localized portion of the graph (e.g., immediate predecessors or successors), making global query evaluation unnecessary. 
As shown in our detailed analysis of query cost in \cref{subsubsec:eval-rq3}, executing the query in small, targeted steps avoids bloated recursive unions, reduces join and memory overhead, and enables early pruning. This step-wise traversal mirrors the investigative workflow, supports effective index reuse, and yields advantages in both performance and semantic alignment.
Our extensive evaluations in \cref{subsubsec:eval-rq2} demonstrate that our scheduler layer consistently improves query performance in different database settings.

\myparatight{In-memory management}
\tool allows users to bind the execution results of a \lang query to a variable, which can then be used as the data source for subsequent queries, avoiding the need to repeatedly fetch data from the database. This in-memory management technique keeps the resulting graph in memory, enabling faster access when the variable is reused. Security analysts can utilize this feature to refine a previously executed query by applying additional constraints, and changes will be applied to the maintained in-memory graph, pruning unnecessary events efficiently.

\section{Evaluation}
\label{sec:eval}

We built \tool ($\sim$15K lines of code) upon several tools:
Sysdig~\cite{sysdig} for Linux system auditing,
Procmon~\cite{procmon} for Windows system auditing,
PostgreSQL~\cite{postgresql} and Neo4j~\cite{neo4j} for data storage, ANTLR 4~\cite{antlr} for the DSL, 
React for the UI,  
and Java for the system.

We conduct extensive evaluations to address key research questions regarding the effectiveness in uncovering attack sequences, query performance, cost, and conciseness, and the usability of our language tool in facilitating iterative attack investigations.

\begin{enumerate}[label={\textbf{(RQ\arabic*)}}, topsep=1pt, partopsep=2pt, leftmargin=*,itemsep=2pt]

    \item How effective is \tool in reducing dependency explosion and uncovering attack sequences in diverse scenarios?

    \item How efficient is \toolno’s scheduler in executing \lang queries over large system provenance data?

    \item How does \toolno's scheduler optimize query cost?

    \item How concise are \lang queries?

    \item How does \langno’s usability compare to general languages?

\end{enumerate}

\subsection{Evaluation Setup}
\label{subsec:eval-setup}

We deployed \tool on a server equipped with an AMD EPYC 7313 (3.00GHz) CPU and 1TB of RAM, running 64-bit Ubuntu 20.04.
We deployed monitoring agents on nine hosts (eight Linux and one Windows machines), which, together with the server, form a controlled local network. All machines are connected through a high-speed Ethernet switch, simulating a typical enterprise network topology with the server acting as a centralized log aggregation and analysis node. Attacks were carried out on these hosts, and the agents collected system events that were transmitted to the server.

We constructed a benchmark of 30 attack cases covering diverse strategies and complexity levels, including 23 executed in our own network and 7 drawn from widely used system audit logs datasets. 
During the attacks, hosts continued to support benign user activities (e.g., file editing, software development), ensuring that realistic background noise was captured alongside malicious activities.

\begin{itemize}[topsep=0pt, partopsep=0pt, leftmargin=*,itemsep=0pt]
    \item \textbf{Multi-step attacks} (8 cases, ``\emph{multistep\_}''): We carried out these attacks across multiple Linux hosts. Each attack begins with an initial penetration from an external host, followed by malware infection, lateral movement across hosts, and data exfiltration. 
    The attacks exploit different vulnerabilities (e.g., Netcat backdoors, Shellshock, and EternalBlue) based on known CVEs~\cite{cve}.

    \item \textbf{Malicious exploits} (10 cases, ``\emph{malicious\_}''): We carried out these attacks on a Linux host based on widely used exploits reported in previous studies~\cite{liu2018priotracker,hassan2019nodoze}.

    \item \textbf{Malware samples} (5 cases, ``\emph{malware\_}''): We selected five malware samples from VirusSign and executed them on our Windows host to infect the system or create a backdoor.

    \item \textbf{DARPA TC cases} (3 cases, ``\emph{tc\_}''): We selected three APT attack cases from the DARPA TC Engagement 5 data release~\cite{darpatc}. These attacks, conducted by a red team in a controlled environment, span multiple hosts and consist of multiple stages. Each case involves three hosts, with system audit logs collected during the engagement to capture both benign and malicious activities.

    \item \textbf{ATLASv2 cases} (4 cases, ``\emph{atlasv2\_}''): 
    We selected four cases from the recent ATLASv2 Attack Engagement dataset~\cite{riddle2023atlasv2atlasattackengagements}, which provides realistic system logs with high-quality background benign noise. Two hosts were involved. 
    Our cases comprise multiple steps and exploit various vulnerabilities in widely used software (e.g., Adobe Flash Player and Microsoft Word).

\end{itemize}

\cref{tab:case-name} lists all 30 cases. The collected logs contain 1,548,202 system entities and 58,388,158 events. Our evaluations span 20 hosts in total: 9 from our own setup, 9 from DARPA TC, and 2 from ATLASv2. Detailed case descriptions are available at~\cite{provexa-website}.

\begin{table}[t]
\caption{30 attack cases in our evaluation benchmark}
\label{tab:case-name}
\begin{adjustbox}{width=\linewidth,center}
\scriptsize
\begin{tabular}{p{0.28\linewidth}p{0.50\linewidth}p{0.1\linewidth}}
\toprule
\textbf{Case ID}         & \textbf{Case Name}   & \textbf{OS}   \\
\midrule
multistep\_cmd\_injection      & Command-line injection    & Linux\\
multistep\_data\_leakage       & Data leakage after penetration & Linux\\
multistep\_netcat\_backdor     & Leaving a netcat backdoor after penetration & Windows\\
multistep\_password\_crack     & Password crack after penetration & Linux\\
multistep\_penetration         & Shellshock penetration & Linux\\
multistep\_phishing\_email     & Phishing email & Windows\\
multistep\_supply\_chain       & Supply chain attack & Linux\\
multistep\_wannacry            & WannaCry attack & Windows\\
malicious\_wget  &  Using wget to download and execute script & Linux\\
malicious\_illegal\_store  & Storing in another user's home directory & Linux\\
malicious\_hide\_file   &  Downloading and hiding malicious file & Linux\\
malicious\_backdoor\_dl  &  Downloading backdoor malware with noise & Linux\\
malicious\_server\_usr  &  Server user performing malicious actions & Linux\\
malicious\_ssh\_theft  & Adding public key to another user's profile  & Linux\\
malicious\_gcc\_crash  & Using gcc to compile and run code that crashes the system.  & Linux \\
malicious\_scan\_login  & Nmap scan and login for ssh & Linux\\
malicious\_pwd\_reuse  &  Password decode with John the Ripper password cracker & Linux\\
malicious\_student &  Student changing midterm score on server & Linux\\
malware\_autorun         & Trojan.Autorun  & Windows\\
malware\_danger          & Trojan.Danger & Windows\\
malware\_hijack          & Virus.Hijack & Windows\\
malware\_infector        & Virus.Infector & Windows\\
malware\_sysbot          & Virus.Sysbot & Windows\\
tc\_fivedirections\_1 & 05092019 1326 - FiveDirections 2 - Firefox Drakon APT Elevate Copykatz Sysinfo    & Linux   \\
tc\_fivedirections\_2 & 05172019 1226 - FiveDirections 3 - Firefox DNS Drakon APT FileFilter-Elevate  & Linux\\
tc\_theia         & 05152019 1448 - THEIA 1 - Firefox Drakon APT BinFmt-Elevate Inject   & Linux   \\

atlasv2_s1     &  Adobe Flash Exploit - CVE-2015-5122  &   Windows \\
atlasv2_s2    &  Adobe Flash Exploit - CVE-2015-3105  &   Windows \\

atlasv2_s3    &  Microsoft Word Exploit - CVE-2017-11882  &   Windows \\
atlasv2_s4     &  Microsoft Word Exploit -  CVE-2017-0199  &  Windows \\

\bottomrule
\end{tabular}
\end{adjustbox}
\end{table}


\subsection{Case Study: APT Attack Investigation}
\label{subsec:query-cases}

We demonstrate how \lang is used to investigate the data leakage case (i.e., multistep\_data\_leakage in \cref{tab:case-name}) described in \cref{sec:background}.
To avoid bias, we assume no prior knowledge of the attack: an author, uninvolved in constructing the attack, independently used \lang to iteratively uncover the attack sequence.

\begin{itemize}[topsep=2pt, partopsep=2pt, listparindent=\parindent, leftmargin=*, itemsep=2pt]

\item \textbf{\emph{Query \circled{1}:}}
We search for a malicious pattern on Host 1, where a sensitive \incode{tar} file is transferred to an unknown IP. The results confirm the presence of this pattern and identify the file as \incode{sensitive_data.tar}, confirming data exfiltration. We bound these events to an in-memory variable \incode{poi1} for further analysis.

\begin{lstlisting}[style=myStyleMain] 
poi1 = search from db(host1) where e1{name="curl", type=process}, e2{name="*.tar", type=file}, e3{type=network} with e2[read]->e1 &&[<1s] e1[write]->e3;
display poi1;
\end{lstlisting}

\item \textbf{\emph{Query \circled{2}:}} We investigate the origin of the \incode{sensitive\_data.tar} file by tracing the dependencies of \incode{poi1} on Host 1, excluding benign entities to narrow the scope.
This backward dependency graph reveals that an \incode{scp} process created this file after reading data from an IP associated with Host 2. This confirms that the attacker used \incode{scp} on Host 1 to transfer the sensitive file from Host 2.

\begin{lstlisting}[style=myStyleMain] 
g1 = back track poi1 from db(host1) exclude nodes where name like "vscode"; display g1;
\end{lstlisting}

\item \textbf{\emph{Query \circled{3}:}}
We further investigate how \incode{sensitive_data.tar} was created on Host 2 by issuing a backward tracing query. The results show that a \incode{tar} process read the password files \incode{/etc/passwd} and \incode{/etc/shadow}, then wrote the data to \incode{sensitive_data.tar}, confirming the sensitive data collection step of the attack.

\begin{lstlisting}[style=myStyleMain]  
g2 = back track "sensitive_data.tar" from db(host2)
    exclude nodes where name like "vscode";
display g2;
\end{lstlisting}

\item \textbf{\emph{Query \circled{4}:}} 
We want to further identify the attack's entry point on Host 1 by tracing the backward dependencies of the malicious \incode{curl} process, while filtering out non-critical processes such as \incode{ping}.
The backward dependency graph is bound to \incode{g3}.

\begin{lstlisting}[style=myStyleMain]  
g3 = back track where exename like "curl" from db(host1) include nodes where not path like "ping";
display g3;
\end{lstlisting}

\item \textbf{\emph{Query \circled{5}:}} Since the attacker is likely operating from a remote server using the same IP (i.e., \incode{20.69.152.188}, as revealed in Query \circled{1}), we search for this IP in \incode{g3} (in-memory). This reveals a \incode{lighttpd} process, which we bind to \incode{poi2}. The results suggest that the attacker exploited a vulnerability in the \incode{lighttpd} web server to initially compromise Host 1 from the attacker’s host \incode{20.69.152.188}.

\begin{lstlisting}[style=myStyleMain] 
poi2 = search from g3 where e1{srcip="20.69.152.188"}, e2{type=process} with e1[read]->e2;
display poi2;
\end{lstlisting}

\item \textbf{\emph{Query \circled{6}:}}
With the entry points identified, we reconstruct the full attack sequence on Host 1. We merge \incode{g1} and \incode{g3} via a union operation, storing the result in \incode{g4}. Forward dependencies from the entry point \incode{poi2} are then traced within \incode{g4} and saved as \incode{g5}. 
Intersecting forward and backward traces helps significantly reduce the backward dependency graph.

\begin{lstlisting}[style=myStyleMain]  
g4 = g1 | g3;
g5 = forward track poi2 from g4 exclude nodes where name like "cat"; display g5;
\end{lstlisting}

\end{itemize}

The dark black paths in \cref{fig:demo} show the results of Query \circled{6}, highlighting the critical attack steps with significantly reduced graph size.
This demonstrates the effectiveness of our \lang in supporting iterative investigation of complex, multi-step attacks.
Following a similar procedure, 
we construct \lang queries for each attack case and use them in evaluations.

\subsection{RQ1: Attack Investigation Effectiveness}
\label{subsubsec:eval-rq1}

We evaluate \toolno's effectiveness in mitigating dependency explosion and uncovering attack sequences, by comparing it with one seminal method (BackTracker~\cite{backtracking}) and two state-of-the-art techniques (Priotracker~\cite{liu2018priotracker} and DepImpact~\cite{fang2022backpropagating}).
BackTracker reconstructs event sequences leading to POIs by tracing dependencies, but suffers from explosion.
PrioTracker improves this by prioritizing abnormal dependencies based on event rareness within a time limit. 
DepImpact assigns weights to edges using features like data flow and temporal relevance, propagates impact scores from POIs to entry nodes, and identifies critical paths by intersecting forward traces from top-ranked entry nodes with backward traces. 
We followed the original parameter settings in these papers and validated our reimplementations with the respective authors.

\myparatight{Mitigating dependency explosion}
As in PrioTracker and DepImpact, we compute the \textbf{\emph{Graph Reduction Rate (GRR)}}, defined as the ratio of edges in the original provenance graph to those in the graph produced by the investigation.
\tool achieves the highest $GRR$ of $58,990.7\times$, significantly outperforming BackTracker $(9.4\times)$, PrioTracker $(41.6\times)$ and DepImpact $(24.3\times)$.
Moreover, the graphs generated by \tool are closest in size to the attack ground truth.
BackTracker produces much larger graphs due to its lack of control. 
PrioTracker performs better than BackTracker in large TC cases due to its prioritization of events.
For small cases, the graph generated by PrioTracker is the same as BackTracker, as the tracking can be finished within the time limit.
However, it is hard to select a proper time limit across scenarios. 
Similarly, DepImpact offers limited user control. Its performance heavily depends on the choice of entry nodes: too few may miss critical paths, while too many introduce noise.
This lack of flexibility limits DepImpact’s effectiveness.

\begin{table}[t]
    \centering
    \caption{Effectiveness in uncovering attack sequences}
    \label{tab:query-acc-averages}
    \centering
\setlength\tabcolsep{3pt}
\scalebox{.85}{
    \begin{tabular}{@{}lccc@{}}
        \toprule
        \textbf{Approach} & \textbf{F1 Score} & \textbf{Miss Ratio (MR)} & \textbf{Noise Ratio (NR)} \\ \midrule
        BackTracker & 0.2526 & 0.1855  & 0.7731 \\
        PrioTracker & 0.2526 & 0.1855 & 0.7731  \\
        DepImpact   & 0.2604 & 0.3999 & 0.6578  \\
        \textbf{\tool}       & \textbf{0.8766} & \textbf{0.0525} & \textbf{0.1658}  \\
        \bottomrule
    \end{tabular}
    }
\end{table}

\myparatight{Uncovering attack sequences}
We use metrics adopted from prior work~\cite{fang2022backpropagating,liu2018priotracker}: \textbf{\emph{miss rate (MR)}}, \textbf{\emph{noise ratio (NR)}}, and \textbf{\emph{F1-score}}.
Miss rate, defined as $MR = FN/E_c$, measures the proportion of attack-relevant events that are not identified.
Noise ratio, defined as $NR = FP/E_{total}$, measures the proportion of irrelevant events included.
F1-score is computed as $F1 = \frac{TP}{(TP + 0.5\times(FP + FN))}$.
$TP$ is true positives.
$FP$ is false positives.
$FN$ is false negatives.
$E_c$ is the attack-relevant event count.
$E_{total}$ is the total event count.

\cref{tab:query-acc-averages} presents the aggregated results. \tool significantly outperforms existing approaches, achieving higher $F1$ and lower $MR$ and $NR$, indicating its effectiveness in accurately revealing attack sequences while filtering out irrelevant events.
BackTracker suffers from high $NR$ due to dependency explosion, producing large graphs (averaging 2,022 nodes and 230,253 edges). It also has a higher $MR$ because it excludes events occurring after the POI in its backward analysis.
PrioTracker shows higher $MR$ and $NR$ than \tool. While it generates smaller graphs than BackTracker in large TC cases (4.42$\times$ fewer edges), the results are still noisy ($NR$ near 1). In smaller cases, it produces identical graphs to BackTracker.
\tool also outperforms DepImpact on both $MR$ and $NR$, benefiting from \langno’s customizable queries. DepImpact’s performance is sensitive to the choice of entry nodes, which is difficult to optimize across cases. For instance, in \emph{multistep_penetration}, it misses many initial access attempts, resulting in a high $MR$ (0.93).
While \toolno’s performance is also influenced by parameter choices, its human-in-the-loop refinement through a DSL enables adaptive, case-specific tuning that reduces noise and improves accuracy.

\subsection{RQ2: Query Execution Efficiency}
\label{subsubsec:eval-rq2}

We evaluate the \textbf{\emph{execution time}} of \tool through two complementary analyses. First, we perform an ablation study to quantify the performance gains from \toolno’s in-memory management. Second, we compare query execution times under different database optimization settings to understand the performance gains from \toolno's domain-aware scheduler.

For each \lang query, we construct semantically equivalent SQL and Cypher queries. As detailed in \cref{sec:execution}, for search queries, we use SQL \incode{CROSS JOIN} and Cypher \incode{MATCH} to specify identical subgraph patterns.
For tracking queries, we use SQL \incode{WITH RECURSIVE} and Cypher’s variable-length path matching with filters.

\subsubsection{Impact of In-Memory Management}
\label{subsubsec:in-memory}

We evaluate \toolno’s runtime performance with and without in-memory management using a benchmark inspired by DepImpact. Each attack case consists of a backward query from POI alerts, a search for key entry nodes, a forward query from those nodes, and an intersection of forward and backward traces to reduce the graph size.
We compare four settings: \emph{provexa-in}, \emph{provexa-db}, \emph{sql}, and \emph{cypher}.
\emph{provexa-in} loads data from the database for the initial backward query and executes subsequent queries in memory. To ensure a fair comparison, we store intermediate results in temporary databases.
\emph{provexa-db} loads data from the original and temporary databases for each query.
SQL and Cypher do not support in-memory execution on prior results, so we run semantically equivalent SQL and Cypher queries on the original and temporary databases (i.e., the same data as \emph{provexa-db}).

We report the total runtime across all queries:
\emph{provexa-in} significantly outperforms \emph{provexa-db}, \emph{sql}, and \emph{cypher}, with $37\times$, $41\times$, $68\times$ speedups, respectively.
This highlights the performance gains of \toolno's in-memory management.

\subsubsection{Impact of Database Optimizations}
\label{subsubsec:db-optimizations}

\begin{table}[t]
\centering
\caption{PostgreSQL optimization configurations}
\label{tab:tuning-settings}
\begin{adjustbox}{width=\linewidth,center}

\begin{tabular}{lll}
\toprule
\textbf{Setting} & \textbf{Categories} & \textbf{Representative Parameters} \\
\midrule
Default &
None &
\scriptsize No changes \\
\midrule
Light &
Planner, memory &
\scriptsize
\begin{tabular}[t]{@{}l@{}}
random\_page\_cost = 1.1\\
work\_mem = 128\,MB\\
effective\_cache\_size = 12\,GB\\
default\_statistics\_target = 200
\end{tabular} \\
\midrule
Moderate &
+ parallelism and collapse &
\scriptsize
\begin{tabular}[t]{@{}l@{}}
\textit{\textbf{Light settings, plus:}}\\
join\_collapse\_limit = 12\\
from\_collapse\_limit = 12\\
parallel\_setup\_cost = 100\\
parallel\_tuple\_cost = 0.01
\end{tabular} \\
\midrule
Advanced &
+ I/O, JIT &
\scriptsize
\begin{tabular}[t]{@{}l@{}}
\textit{\textbf{Moderate settings, plus:}}\\
shared\_buffers = 4\,GB\\
temp\_buffers = 64\,MB\\
jit = on\\
autovacuum\_vacuum\_scale\_factor = 0.1
\end{tabular} \\
\midrule
\textbf{Final} & Planner, memory, collapse &
\scriptsize
\begin{tabular}[t]{@{}l@{}}
random\_page\_cost = 1.1\\
work\_mem = 64\,MB\\
effective\_cache\_size = 8\,GB\\
join\_collapse\_limit = 8
\end{tabular} \\
\bottomrule
\end{tabular}

\end{adjustbox}
\end{table}

PostgreSQL's default configuration is conservative, prioritizing stability and broad compatibility across low-resource environments. While not optimized for performance, we include this \emph{Default} setup as a baseline to measure tuning benefits. In contrast, \langno’s tracking and search queries are memory-intensive and involve complex joins. To assess how tuning alone can affect performance, we tested three additional configurations--\emph{Light}, \emph{Moderate}, \emph{Advanced}--as shown in Table~\ref{tab:tuning-settings}. These incrementally adjust memory parameters, planner cost settings, join-collapsing thresholds, parallelism, and I/O behaviors.

We evaluated these configurations with our benchmark queries. The results show that while each tuning level offered incremental gains over the baseline, improvements were modest and inconsistent across workloads. This variability reflects the structural diversity of our benchmark queries. No single configuration consistently dominated. Therefore, we combined parameters to derive a configuration that achieved the lowest average query cost and execution time across all queries, denoted as the \emph{Final} configuration. Detailed results for all tuning levels are available at~\cite{provexa-website}.


\subsubsection{Impact of Domain-Aware Scheduler}
\label{subsubsec:query-performance}

\begin{table*}[t]
\caption{Execution time (in milliseconds) for PostgreSQL and Neo4j with various optimization settings, both with and without \toolno's scheduler. 
Each query is executed for 10 rounds. The time-out threshold is 30 minutes.
We exclude DARPA TC and ATLASv2 cases (marked with "-") from the Neo4j database evaluation due to their time-out in data loading. \textbf{V} - Vanilla DB, \textbf{O} - Optimized version of DB, \textbf{VR} - Vanilla with \toolno's scheduler, \textbf{OR} - Optimized version of DB with \toolno's scheduler.}

\begin{adjustbox}{width=.95\linewidth,center}
\begin{tabular}{lcccc|cccc|cccc|cccc}
\hline
& \multicolumn{8}{c}{\textbf{Tracking}} 
& \multicolumn{8}{c}{\textbf{Search}} \\
\cline{2-17}

& \multicolumn{4}{c}{\textbf{PostgreSQL}} 
& \multicolumn{4}{c}{\textbf{Neo4j}} 
& \multicolumn{4}{c}{\textbf{PostgreSQL}} 
& \multicolumn{4}{c}{\textbf{Neo4j}} \\
\cline{2-17}
\multirow{-3}{*}{\textbf{Case}} 
& V & O & VR & OR & V & O & VR & OR & V & O & VR & OR & V & O & VR & OR \\ 
\hline
\textbf{multistep\_cmd\_injection}  & 176 & 236 & 2 & 2 & 2,879 & 2,538 & 852 & 769 & 46 & 50 & 52 & 56 & 671 & 576 & 579 & 569 \\

\textbf{multistep\_data\_leakage}   & 684 & 631 & 6 & 3 & 18,781 & 18,104 & 2093 & 1733 & 323 & 265 & 301 & 286 & 802 & 701 & 795 & 694 \\

\textbf{multistep\_netcat\_backdor} & 524 & 570 & 5 & 5 & 28,513 & 28,247 & 3,323 & 2,898 & 17 & 17 & 18 & 18 & 473 & 400 & 448 & 414 \\

\textbf{multistep\_password\_crack} & 232 & 312 & 5 & 4 & 4,649 & 4,542 & 1,636 & 1,374 & 86 & 80 & 97 & 94 & 711 & 651 & 671 & 647 \\

\textbf{multistep\_penetration} & 1,011 & 922 & 2 & 1 & 2,054 & 1,672 & 689 & 674 & 488 & 429 & 531 & 447 & 720 & 643 & 765 & 665 
\\ 

\textbf{multistep\_phishing\_email} & 6,394 & 5,750 & 10 & 10 & Time-out & Time-out & 3,558 & 3,101 & 1,996 & 1,693 & 68 & 68 & Time-out & Time-out & 559 & 467
\\

\textbf{multistep\_supply\_chain} & 22 & 112 & 1 & 1 & 2,737 & 2,430 & 500 & 487 & 10 & 9 & 9 & 10 & 463 & 434 & 487 & 433
\\

\textbf{multistep\_wannacry} & 318 & 416 & 1 & 2 & 1,248,315 & 1,301,313 & 2,194 & 1,904 & 159 & 161 & 157 & 166 & 506 & 457 & 631 & 450
\\

\textbf{malicious\_wget} & 495 & 513 & 2 & 1 & 5,494 & 5,190 & 775 & 609 & 27,490 & 21,565 & 173 & 164 & 739 & 617 & 626 & 610
\\

\textbf{malicious\_illegal\_store} & 287 & 349 & 2 & 1 & 4,424 & 4,101 & 826 & 729 & 146 & 155 & 150 & 171 & 544 & 452 & 489 & 447
\\

\textbf{malicious\_hide\_file} & 515 & 528 & 1 & 1 & 2,224 & 1,965 & 675 & 522 & 139 & 150 & 144 & 159 & 428 & 443 & 489 & 452
\\

\textbf{malicious\_backdoor\_bl} & 558 & 564 & 2 & 1 & 5,577 & 5,062 & 1,038 & 857 & 149 & 150 & 156 & 162 & 554 & 443 & 517 & 443
\\

\textbf{malicious\_server\_usr} & 492 & 462 & 3 & 1 & 13,843 & 13,408 & 961 & 883 & 2,530 & 1,967 & 659 & 594 & 891 & 825 & 944 & 860
\\

\textbf{malicious\_ssh\_theft} & 287 & 323 & 1 & 1 & 2,166 & 1,893 & 810 & 687 & 130 & 104 & 130 & 112 & 737 & 717 & 836 & 683
\\

\textbf{malicious\_gcc\_crash} & 338 & 336 & 6 & 5 & 3,039 & 2,807 & 964 & 801 & 139 & 108 & 135 & 124 & 819 & 683 & 736 & 677
\\

\textbf{malicious\_scan\_login} & 77 & 166 & 4 & 3 & 738,173 & 713,940 & 1,230 & 1,081 & 8,682 & 7,287 & 10,602 & 9,354 & 686 & 617 & 719 & 616
\\

\textbf{malicious\_pwd\_reuse} & 78 & 173 & 5 & 5 & 54,134 & 54,025 & 786 & 667 & 62 & 64 & 56 & 65 & 509 & 461 & 504 & 437
\\

\textbf{malicious\_student} & 12 & 105 & 3 & 2 & 1,801 & 1,520 & 908 & 771 & 10 & 11 & 11 & 11 & 526 & 431 & 527 & 439
\\

\textbf{malware\_autorun} & 2,682 & 2,544 & 5 & 5 & 23,004 & 21,658 & 1,707 & 1,545 & 29 & 31 & 30 & 33 & 464 & 435 & 488 & 434
\\

\textbf{malware\_danger} & 201,698 & 186,305 & 11 & 12 & 1,006,627 & 1,037,522 & 16,302 & 15,355 & 124 & 642 & 120 & 667 & 707 & 565 & 725 & 571
\\

\textbf{malware\_hijack} & 112,522 & 102,836 & 18 & 11 & 1,141,741 & 1,162,804 & 6,983 & 6,395 & 154 & 147 & 160 & 156 & 592 & 537 & 682 & 523
\\

\textbf{malware\_infector} & 10,268 & 9,569 & 9 & 9 & 545,064 & 536,059 & 3,231 & 2,761 & 102 & 96 & 77 & 69 & 621 & 599 & 724 & 587
\\

\textbf{malware\_sysbot} & 5,958 & 5,367 & 9 & 8 & 181,274 & 277,693 & 2,958 & 2,546 & 40 & 42 & 47 & 46 & 594 & 450 & 527 & 440
\\

\textbf{tc\_fivedirections\_1} & 113,757 & 113,140 & 98 & 93 & - & - & - & - & 28,699 & 40,467 & 27,667 & 41,124 & - & - & - & - 
\\

\textbf{tc\_fivedirections\_2} & 128,568 & 134,492 & 304 & 264 & - & - & - & - & 52,200 & 68,633 & 41,702 & 42,454 & - & - & - & -
\\

\textbf{tc\_theia} & 75,956 & 71,837 & 301 & 245 & - & - & - & - & 11,682 & 14,216 & 11,307 & 14,577 & - & - & - & -
\\

\textbf{atlasv2_s1}   & 3,129  &  2,569 & 2  & 1  & - & - & - & - &  1,507 &  1,106 & 1,487  &  1,070  & - & - & - & -   \\

\textbf{atlasv2_s2}   & 3,531  &  3,110 & 2  &  1 & - & - & - & - &  1,182 &  909 &  1,198 &  840 & - & - & - & -   \\

\textbf{atlasv2_s3}   & 2,796  & 2,653  & 1 & 1  & - & - & - & - & 1,195  &  926 & 1,218  &  988  & - & - & - & -   \\

\textbf{atlasv2_s4}   & 1,565  &  1,412 & 3  &  3 & - & - & - & - &  1,233 & 931  & 1,091  &  919  & - & - & - & -    \\

\hline
\textbf{Average}  & 22,498 & 21,610 & 27 & 24 & 228,932 & 236,295 & 2,391 & 2,137 & 4,692 & 5,414 & 3,319 & 3,833 & 625 & 551 & 629 & 546 \\
\hline
\end{tabular}
\end{adjustbox}
\label{tab:exec-time-psql-full}
\end{table*}

We evaluate the execution time gains from \toolno’s scheduler layer across four configurations: \emph{vanilla}, \emph{optimized}, \emph{vanilla-provexa}, and \emph{optimized-provexa}.
\emph{vanilla} and \emph{optimized} correspond to the \emph{Default} and \emph{Final} database configurations described in \cref{subsubsec:db-optimizations}.
\emph{vanilla} reflects the out-of-the-box setup commonly seen in many deployments, while \emph{optimized} represents a best-effort tuning. 
\emph{vanilla-provexa} layers \toolno’s domain-aware scheduler on top of the default engine, and \emph{optimized-provexa} combines the scheduler with the tuned engine.
We followed a similar tuning approach to Neo4j.
\lang queries are executed in the \emph{vanilla-provexa} and \emph{optimized-provexa} setups, whereas semantically equivalent SQL and Cypher queries are used for \emph{vanilla} and \emph{optimized}. 
Each query is executed 10 times per configuration, with caches cleared and the databases restarted between runs to ensure fairness and eliminate residual effects.

Table~\ref{tab:exec-time-psql-full} shows the results. 
\toolno’s scheduler consistently improves performance across all database configurations (comparing \emph{vanilla-provexa} to \emph{vanilla}, and \emph{optimized-provexa} to \emph{optimized}).
For tracking queries, optimized PostgreSQL outperforms the vanilla setup, while Neo4j’s tuning yields mixed results (improving in some cases, degrading in others). 
This highlights the inherent difficulty of crafting tuning strategies that generalize across all attack cases.
In contrast, the integration of \toolno's domain-aware scheduling yields significant speedups, reaching up to 900$\times$ in PostgreSQL and 110$\times$ in Neo4j when layered atop optimized engines.

For search queries, improvements are more modest. In PostgreSQL, the optimized setup slightly underperforms the vanilla version, further illustrating the challenge of identifying optimization parameters that consistently improve performance across diverse workloads. 
The introduction of \tool nonetheless results in measurable gains--1.43$\times$ in PostgreSQL and 1.01$\times$ in Neo4j.
These modest gains stem from the simplicity of some search queries, where native planners already perform well and the benefits of additional processing by \toolno's scheduler are less pronounced.


\myparatight{DuckDB database}
We extend evaluation to DuckDB~\cite{duckdb}, a fast, in-process analytical database for single-node OLAP workloads. Unlike PostgreSQL and Neo4j, DuckDB offers a lightweight, embedded alternative.
To assess \toolno’s scheduler on DuckDB, we compare vanilla SQL against \lang queries executed using the scheduler.
Queries ran on DuckDB’s native file format, leveraging its columnar storage and vectorized execution engine. Caching was disabled, and results were averaged over 10 trials.
The results show that \tool delivers substantial performance gains for tracking queries and complex search queries in DuckDB (averaging 148ms vs. 1.4s with vanilla SQL, a 9.4$\times$ speedup), mirroring improvements seen on other backends.
For simpler search queries, \toolno's scheduler provides less speedup, as direct SQL execution is already efficient.

Despite its fast performance, DuckDB exhibits limitations under investigative workloads involving large provenance data.
In such settings, particularly during evaluation of DARPA TC scenarios (e.g., \textit{tc\_fivedirections\_1}, including 225K nodes and 23M edges), DuckDB frequently ran out of memory and crashed during tracking queries. In contrast, PostgreSQL remained stable and completed the same queries, albeit with higher execution times.
Notably, \toolno's scheduler reduced PostgreSQL's runtime from 113 seconds to 93 milliseconds for this case, demonstrating both its scalability and resilience under heavier workloads.

\emph{\textbf{Summary:}} 
The results show that across all cases, query types, database systems (PostgreSQL, Neo4j, DuckDB), and configuration settings (vanilla, optimized), adding \toolno’s scheduling layer consistently improves performance.
These gains are particularly evident for complex queries, where conventional database tuning often yields limited or inconsistent benefits. 
In contrast, \tool provides reliable gains through a system provenance–aware decomposition and execution strategy that prioritizes high-pruning, semantically meaningful query steps. 
Overall, these results highlight \toolno’s generalizability and robustness as a complementary optimization layer tailored to the system provenance domain.

\subsection{RQ3: Cost Estimation}
\label{subsubsec:eval-rq3}

\begin{table*}[t]
\caption{Cost estimation of PostgreSQL with and without \toolno's scheduler. Each query is executed for 10 rounds.}

\begin{adjustbox}{width=.95\linewidth,center}
\begin{tabular}{lcccc|cccc}
\hline
& \multicolumn{4}{c}{\textbf{Tracking}} 
& \multicolumn{4}{c}{\textbf{Search}} \\
\cline{2-9}

\cline{2-9}
\multirow{-3}{*}{\textbf{Case}} 
& Vanilla & Optimized & Vanilla-Provexa & Optimized-Provexa& Vanilla & Optimized & Vanilla-Provexa & Optimized-Provexa \\ 
\hline
\textbf{multistep\_cmd\_injection} & 1.7E+14 & 6.1E+13 & 436 & 169 & 3,225 & 2,524 & 3,138 & 2,436 \\

\textbf{multistep\_data\_leakage} & 8.3E+18 & 2.8E+18 & 1,315 & 507 & 38,978 & 10,782 & 38,765 & 10,568 \\

\textbf{multistep\_netcat\_backdor} & 3.3E+8 & 1E+8 & 486 & 258 & 902 & 725 & 873 & 696 \\

\textbf{multistep\_password\_crack} & 2E+16 & 7.7E+15 & 1,096 & 419 & 5,225 & 3,938 & 5,139 & 3,851 \\

\textbf{multistep\_penetration} & 1E+20 & 3.9E+19 & 1,045 & 375 & 71,395 & 19,587 & 71,004 & 19,195 \\

\textbf{multistep\_phishing\_email} & 1.6E+11 & 6E+10 & 1,389 & 769 & 13,074 & 4,504 & 10,455 & 2,958 \\

\textbf{multistep\_supply\_chain} & 7.4E+7 & 2.4E+7 & 163 & 102 & 596 & 464 & 582 & 450 \\

\textbf{multistep\_wannacry} & 3.7E+6 & 2E+6 & 93 & 71 & 354 & 285 & 343 & 274 \\

\textbf{malicious\_backdoor\_bl} & 2.2E+15 & 7.2E+14 & 251 & 98 & 24,142 & 6,665 & 23,863 & 6,386 \\

\textbf{malicious\_gcc\_crash} & 5.9E+14 & 2.2E+14 & 710 & 523 & 18,197 & 4,988 & 18,079 & 4,870 \\

\textbf{malicious\_hide\_file} & 2.2E+15 & 7.4E+14 & 167 & 58 & 24,124 & 6,663 & 23,844 & 6,384 \\

\textbf{malicious\_illegal\_store} & 2.4E+15 & 8.1E+14 & 134 & 46 & 24,693 & 6,900 & 24,410 & 6,617 \\

\textbf{malicious\_pwd\_reuse} & 4.8E+11 & 1.67E+11 & 391 & 271 & 3,456 & 2,692 & 3,376 & 2,612 \\

\textbf{malicious\_scan\_login} & 6E+11 & 2E+11 & 352 & 244 & 5,634 & 3,968 & 3,481 & 2,693 \\

\textbf{malicious\_server\_usr} & 1.3E+16 & 5.5E+15 & 552 & 190 & 46,488 & 14,832 & 452,117 & 269,703 \\

\textbf{malicious\_ssh\_theft} & 7.5E+16 & 2.9E+16 & 148 & 53 & 18,351 & 5,010 & 18,233 & 4,891 \\

\textbf{malicious\_student} & 1.3E+6 & 3.5E+5 & 214 & 76 & 258 & 204 & 252 & 197 \\

\textbf{malicious\_wget} & 1.2E+15 & 4.2E+14 & 225 & 85 & 28,900 & 9,550 & 24,043 & 6,815 \\

\textbf{malware\_autorun} & 7.8E+9 & 3.8E+9 & 668 & 367 & 1,998 & 1,601 & 1,936 & 1,539 \\

\textbf{malware\_danger} & 1.9E+12 & 6.7E+11 & 2,017 & 1,149 & 22,319 & 6,108 & 22,057 & 5,846 \\

\textbf{malware\_hijack} & 1.5E+14 & 5.5E+13 & 2,826 & 1,220 & 22,571 & 6,817 & 22,320 & 6,566 \\

\textbf{malware\_infector} & 2.1E+10 & 7.8E+9 & 1,361 & 699 & 4,955 & 3,526 & 3,331 & 2,576 \\

\textbf{malware\_sysbot} & 9E+9 & 4.4E+9 & 1,260 & 648 & 2,596 & 2,092 & 2,511 & 2,006 \\

\textbf{tc\_fivedirections\_1} & 3.3E+23 & 1.3E+23 & 61,785 & 25,527 & 8.9E+6 & 3.4E+6 & 8.8E+6 & 3.3E+6 \\

\textbf{tc\_fivedirections\_2} & 1.5E+12 & 5.1E+11 & 148,555 & 60,180 & 1.2E+7 & 5.1E+6 & 1.1E+7 & 3.9E+6 \\

\textbf{tc\_theia} & 1.5E+11 & 5.7E+10 & 136,603 & 49,401 & 2.4E+6 & 1.2E+6 & 2.3E+6 & 1.2E+6 \\

\textbf{atlasv2_s1}  & 1.3E+9  &  4E+8 &  285  &  98  & 169,574  &  37,084  &  168,362  &   35,871 \\

\textbf{atlasv2_s2}  &  2.6E+21 & 9.9E+20  &  253  & 85  &  168,665 &  36,924  &  167,459  &  35,717  \\

\textbf{atlasv2_s3}  &  2.7E+21 & 1E+21  &  361  & 133  &  169,837 &  37,124  &  168,622  &  35,908  \\

\textbf{atlasv2_s4}  &  6.4E+20 &  2.4E+20 &  725  &  254 & 174,938  &  42,181  &  173,731  &  40,973  \\

\hline
\textbf{Average}  & \textbf{1.1E+22} & \textbf{4.3E+21} & \textbf{12,196} & \textbf{4,791} & \textbf{795,132} & \textbf{331,835} & \textbf{773,928} & \textbf{298,299} \\
\hline
\end{tabular}
\end{adjustbox}
\label{tab:cost-estimate-psql-full}
\end{table*}

To further assess the improvements introduced by \toolno’s scheduler upon the database planner, we analyze \textbf{\emph{cost estimates}} across the same four configurations from~\cref{subsubsec:query-performance}: \emph{vanilla}, \emph{optimized}, \emph{vanilla-provexa}, and \emph{optimized-provexa}.
We use \incode{EXPLAIN ANALYZE} to extract PostgreSQL’s internal cost estimates, which approximate expected resource consumption--lower values indicate more efficient plans.
As Neo4j’s PROFILE command does not expose comparable cost metrics, it is excluded from this experiment.

Table~\ref{tab:cost-estimate-psql-full} presents the cost estimates. While manual tuning of PostgreSQL yields modest performance gains, \toolno’s scheduler achieves significantly greater improvements. Across both query types, \tool consistently lowers cost estimates relative to the vanilla setup, with stronger effects when paired with the optimized configuration. The most substantial gains occur when \tool is layered atop the tuned PostgreSQL engine, where its decomposition and scheduling strategies complement the tuned planner.

To better understand why our multi-query execution strategy leads to lower planner-estimated costs, we analyze representative tracking and searching queries. 
As an example, consider the tracking query for the \textit{multistep_supply_chain} case shown below.

\begin{lstlisting}[style=myStyleMain, label={lst:sample-tracking-query}]
bg = back track where (cmdline like "freemem.sh", type=process) from db(case2_supply_chain); display bg;
\end{lstlisting}

This query looks for processes whose command line contains \incode{freemem.sh} and tracks all events that causally precede them, in the \incode{case2_supply_chain} database. 
Without our scheduler, this requires a recursive SQL query that generates a monolithic execution plan processing over 144 million intermediate rows, with an estimated cost of approximately 74 million. 
In contrast, our approach starts with a simple query that selects processes matching \incode{cmdline LIKE 'freemem.sh'} and then iteratively retrieves parent processes through separate follow-up queries. Each step operates over a narrow set of candidates (3 to 140 PIDs), with individual query costs between 22 and 73, adding up to just 163. This strategy avoids the overhead of unconstrained recursive joins and large materialized intermediaries.

A similar benefit appears for search queries. Consider the query for the \textit{malware_infector} case shown below.

\begin{lstlisting}[style=myStyleMain, label={lst:sample-searching-query}]
poi = search from bg where e1{path like "Virus.Infector", type=file}, e2{pid=2688, exename like "exe", type=process}, e3{type=file} with e1->e2 &&[<100s] e2->e3; display poi;
\end{lstlisting}

The goal is to find a file named \incode{Virus.Infector} read by a process with \incode{pid=2688} and exename extension `exe', which then writes to another file within 100 seconds.
This query runs over a prior result \incode{bg} from a backtracking query.
Instead of executing the entire pattern in a single query with multiple joins and constraints, we split it into two subqueries: the first computes \incode{e1->e2} to find file-to-process reads, and the second computes \incode{e2->e3} to find subsequent writes from those processes, applying the temporal constraint (\incode{<100s}) in the second step.
This decomposition reduces the planner-estimated cost by 1.5$\times$ compared to the monolithic version.

\textbf{\emph{Summary:}}
These improvements reflect a key planning advantage: query optimizers are more effective at handling small, localized subqueries than deep recursive plans or multi-join blocks. 
By isolating semantically meaningful constraints early and narrowing each step’s scope, our strategy enables selective index scans and faster execution. 
This approach is particularly effective when queries contain selective constraints that sharply reduce the search space, for example, filtering by a specific script name, IP address, or file path. Such constraints are common in real investigative workflows, where analysts typically start from a point of interest and explore outward.
In cases where constraints are overly general (e.g., searching for any process with \incode{cmdline LIKE 'exe'}), the initial query may return a large result set, reducing the pruning benefits of decomposition. 
Nevertheless, because real-world investigations often start with targeted hypotheses or known indicators, our multi-query decomposition strategy remains well-suited in practice and consistently delivers lower cost and better performance.
By decomposing queries in the application layer while accounting for domain semantics, we play to the strengths of the database engine, which can efficiently execute simple, targeted queries using indexes and optimized access paths. 
The database still applies its own planner optimizations, but we effectively reduce its burden by structuring the workload to align with its strengths.
Detailed query splitting examples and their cost analyses are available at~\cite{provexa-website}.

\subsection{RQ4: Query Length}
\label{subsubsec:eval-rq4}

We compare the \textbf{\emph{query length}} of \lang with SQL, Cypher, and Splunk SPL, an industry-standard security DSL.
Query length is a standard metric in DSL research, as demonstrated in previous works \cite{kosardslgpl,barisicusabilitydsl}, where shorter queries have been shown to improve efficiency and reduce cognitive overhead. 
This metric highlights how \lang simplifies query formulation, reduces verbosity and cognitive load, and facilitates easier iterative attack investigations.
Splunk~\cite{splunk}, a widely used security log analysis solution, employs its own DSL, Splunk SPL, for investigating system logs.
However, Splunk SPL lacks native support for recursive queries needed for tracking~\cite{splunkrecursive}, so our comparison is limited to search queries. 

Example queries are available at~\cite{provexa-website}.
\lang tracking queries are more concise, averaging 19 words vs. 421 in SQL and 183 in Cypher.
While Cypher is more concise than SQL due to its direct modeling of graph, it becomes verbose when using multiple FOREACH loops for traversal.
\lang also leads in search query conciseness, averaging 15 words, compared to 192 in SQL, 25 in Cypher, and 62 in Splunk SPL.
SQL and Splunk SPL require multiple joins to express entity relationships, making complex queries verbose. Cypher, despite modeling relationships directly, still needs verbose MATCH and WHERE clauses for complex constraints, resulting in lengthy queries for expressing complex relationships.

\subsection{RQ5: User Study}
\label{subsubsec:eval-rq5}

We further conducted a user study to evaluate \toolno’s usability and adoption. The study compared \lang with SQL and Cypher, two widely used query languages, to assess efficiency, learning curve, and practicality for attack investigation.
We recruited 10 participants with SQL and Cypher experience, and a security background; several also had data science expertise and familiarity with exploratory analysis, providing broader perspectives.
Each participant completed six investigation tasks using SQL, Cypher, and \lang, divided into two complexity levels. \emph{Simple tasks} involved single-host attacks (\emph{malicious_ssh_theft}, \emph{malicious_illegal_store}, \emph{malicious_server_usr}), while \emph{complex tasks} spanned multiple hosts and required more investigative queries (\emph{multistep_supply_chain}, \emph{multistep_penetration}, \emph{multistep_password_crack}). Task difficulty was balanced within each category.
Participants used SQL for \emph{malicious_ssh_theft} and \emph{multistep_supply_chain}, Cypher for \emph{malicious_server_usr} and \emph{multistep_password_crack}, and \lang for \emph{malicious_illegal_store} and \emph{multistep_penetration}.

To evaluate ease of transitioning between languages, we introduced a transition task focused on complex cases. Participants reattempted tasks using a different language, switching from \lang to SQL/Cypher or vice versa. This design tests whether \langno’s high-level constructs simplify query formulation and reduce cognitive load. Each participant completed four additional transition tasks: \emph{multistep_supply_chain} and \emph{multistep_password_crack} with \lang, and \emph{multistep_penetration} with SQL and Cypher.

Each participant signed a consent form (IRB-approved) and received PostgreSQL and Neo4j databases with provenance data, schema details, and case descriptions outlining investigation goals (e.g., identifying attacker IPs and key PIDs). Simple tasks had a 30-minute limit; complex ones, 45 minutes. Studies were conducted individually on our laptops using \toolno’s UI for \lang, pgAdmin for SQL, and the Neo4j Dashboard for Cypher. 
Screen activity (no audio/video) was recorded to track completion time and query count. Afterward, participants completed a QuestionPro survey rating each language on ease of use, syntax simplicity, and overall experience, using a 5-point Likert scale.

\myparatight{User study results}
For simple cases, \emph{malicious_ssh_theft} using SQL averaged 19 minutes and 3 queries; \emph{malicious_server_usr} using Cypher took 16 minutes and 3.8 queries. In contrast, \emph{malicious_illegal_store} using \lang was completed in just 7 minutes with 1.7 queries.
For complex cases, \lang showed clear efficiency gains. In \emph{multistep_supply_chain}, users improved from 6 minutes with SQL to 4 minutes using \lang. In \emph{multistep_password_crack}, Cypher users averaged 14 minutes and 4.8 queries, but only 6 minutes with \lang.
In \emph{multistep_penetration}, \lang users completed the task in 16 minutes with 3 queries.
By comparison, only 33\% of SQL users finished within the 45-minute limit (average 30 minutes, 5.6 queries). 
30\% of Cypher users timed out, and the rest averaged 12 minutes and 3.8 queries. 
Despite familiarity, SQL’s verbosity led to longer times and more queries, while Cypher showed inconsistent performance. In contrast, \lang was more reliable, requiring fewer queries, and delivering steady performance.

We also analyzed post-survey feedback; detailed questions and results are available at~\cite{provexa-website}.
Participants rated four metrics (ease of query writing, result interpretation, event search, and tracking sequences) on a 5-point scale (Very Complex, Complex, Neutral, Easy, and Very Easy).
\lang consistently received higher ratings, with the most rating it ``Easy'' or ``Very Easy'' across all metrics. Specifically, 100\% rating it “Easy” or better for event search, and 70\% rating it “Very Easy” for tracking.
In contrast, SQL and Cypher received mixed feedback, with many users marking them as ``Complex''or ``Very Complex''. Overall, 95\% of participants found \lang easier to use than SQL or Cypher.
For language transitions, 90\% rated switching from SQL to \lang as ``Easy" or better, while 70\% found the reverse ``Complex" or ``Very Complex''. 
A similar pattern emerged for Cypher.
80\% preferred \lang for attack investigations, and 80\% rated transitioning to it as “Easy” or better. However, returning to Cypher was harder; 40\% rated it ``Complex" and 40\% ``Neutral".
These findings confirm \langno’s usability advantages to streamline attack investigations and reduce cognitive load.

\section{Discussion}
\label{sec:discussion}

\myparatight{Query optimizations}
\tool can benefit from built-in database query optimization strategies while adding its own domain-specific enhancements.
Traditional cost-based optimization selects query plans based on estimated resource costs (e.g., CPU, I/O, memory). \tool provides a domain-aware scheduler that decomposes complex provenance queries based on pruning power, ensuring that high-impact constraints are applied early to reduce the search space.
\toolno's scheduler operates as an additional domain-aware optimization layer atop the database, dispatching its decomposed subqueries to the database’s native scheduler. 
This dual-layer optimization ensures that \tool benefits from both its own optimization strategies and the database's native performance-enhancing techniques.
Future work could explore integrating cost-based decision-making into the decomposition process and extending lazy evaluation to operate across decomposed subqueries.

\myparatight{Data science notebooks}
While computation notebooks like Jupyter are excellent for general-purpose exploration, they are not ideally suited for system provenance analysis in security contexts. Audit logs
are massive in volume, span large time windows, and are stored in databases to support scalable, persistent querying. Notebooks would still require integration with these backends, introducing overhead and setup complexity. Raw logs must be parsed and transformed into structured provenance graphs for identifying complex patterns--capabilities that \tool provides natively.
Nevertheless, \tool offers a notebook-like UI featuring code cells to support iterative exploration.

\section{Related Work}
\label{sec:literature}

\myparatight{Attack investigation using audit logs}
System audit logs are vital resources for investigating complex attacks. Causal dependency tracking methods trace root causes. To mitigate dependency explosion, some works use heuristics to prioritize dependencies~\cite{liu2018priotracker,fang2022backpropagating,hassan2019nodoze}, but risk information loss. Others partition process execution into finer units~\cite{kwon2018mci,ma2016protracer}, but require intrusive instrumentation and kernel modifications, limiting their practical adoption. 
Another class of techniques uses subgraph matching over provenance graphs. 
Poirot~\cite{milajerdi2019poirot} uses heuristics to score node and path alignments, but subsequent studies~\cite{altinisik2023provg,wei2021deephunter,gao2021enabling} report high computational costs and missed attack steps due to its early termination at the first valid alignment. 
Learning-based methods like ProvG-Searcher~\cite{altinisik2023provg} and DeepHunter~\cite{wei2021deephunter} embed graphs using neural models but only determine whether a query graph is entailed, without reconstructing attack steps. They also require costly offline training.

\tool takes a different path. Emphasizing human-in-the-loop analysis, it offers a DSL for flexible, iterative investigation. Unlike prior one-shot approaches, \tool supports incremental refinement and makes analyst intent explicit through DSL query construction. It avoids early stopping, needs no offline training, and uniquely supports native dependency tracking--capabilities missing from prior subgraph matching approaches.

\myparatight{Database query languages}
Existing query languages like SQL and Cypher are not designed for attack investigation. They lack native support for causality tracking, which is essential for tracing multi-step attacks, and their general-purpose execution models perform poorly on system provenance workloads. Queries in these languages often become verbose and error-prone when investigating complex attacks. 
SPARQL~\cite{sparql} and MongoDB’s JSON-based query language~\cite{bradshaw2019mongodb} share similar limitations: SPARQL adopts SQL-like syntax for RDF data, and MongoDB’s model is tailored for document retrieval. Neither supports iterative workflows, efficient modeling of multi-stage attacks, or variable binding across queries, making them inefficient for complex, evolving investigations.
\tool addresses these limitations through a domain-specific abstraction tailored to system provenance and attack investigation. Its DSL enables intuitive event pattern search, chaining, and iterative human-in-the-loop workflows. A domain-aware scheduler leverages event dependencies and pruning to optimize execution, adding a layer atop existing backends for performance enhancement. 
Queries in \lang are more concise, making it easier to incorporate analyst knowledge and refine iteratively. These design choices enable \tool to deliver faster, more intuitive, and more robust forensic analysis than traditional query languages.

\myparatight{Security analysis languages}
Splunk~\cite{splunk} and Elasticsearch~\cite{elasticsearch} are among the most widely used log analysis tools for system performance monitoring and security diagnostics. 
Splunk offers a Search Processing Language (SPL) that combines keyword and regex-based search with shell-like piping for extracting insights from logs. Elasticsearch provides a JSON-based DSL for text search and filtering. 
OSQuery~\cite{osquery}, an open-source tool, uses a SQL-like syntax to query the operating system state.
Unlike \tool, these languages do not support complex dependency tracking queries, which are essential for tracing multi-step attack event chains.
Other specialized languages have been developed for different security tasks, such as network intrusion detection~\cite{chimera, lambda}, secure overlay networks~\cite{mace, networklang}, and threat intelligence representation~\cite{cybox, stix}.
Each is tailored to its own domain and use case.
In contrast, \lang is specially designed for the iterative investigation of complex, multi-stage attacks over system provenance data.
In addition, the two query syntaxes supported by \lang cannot be easily expressed in these languages.

\myparatight{Future directions}
Several future directions can be explored. First, while our current approach prioritizes split queries based on constraints, this can sometimes lead to high intermediate costs when that constraint is not selective. In such cases, a monolithic query planned holistically by the database might be more efficient. 
One extension is to add a lightweight cost-based fallback mechanism that compares the estimated cost of a monolithic plan with the split plan and selects the cheaper option dynamically.
Second, although we apply best-effort tuning to the underlying database engine, \tool achieves the largest gains when paired with an optimized configuration. Since \toolno’s decomposition is orthogonal to engine-level tuning, future work could explore how more adaptive database optimizations further amplify its benefits.
Third, while \tool largely reduces the effort of query formulation compared to general-purpose languages, analysts still have to write queries manually. Leveraging cyber threat intelligence~\cite{McMillan2013,gao2021security,gao2023threatkg,cheng2025ctinexus}, which captures common attack patterns and indicators, along with large language models offers a potential path toward automatically generating \lang queries aligned with investigative goals.

\section{Conclusion}

We presented \tool, a defense system that empowers human analysts to effectively analyze large-scale system provenance to investigate complex multi-step attacks.  \tool introduces an intuitive and expressive domain-specific language, \lang, for capturing diverse attack patterns and analytical workflows. It also features an efficient, domain-aware execution engine that enhances query performance across different database backends.

\begin{acks}
We would like to thank the anonymous reviewers and our shepherd for their constructive feedback. This work is supported in part by the Commonwealth Cyber Initiative (CCI). Any opinions, findings, and conclusions made in this paper are those of the authors and do not necessarily reflect the views of the funding agencies.
\end{acks}


\bibliographystyle{ACM-Reference-Format}
\bibliography{references}


\begin{thebibliography}{54}


\ifx \showCODEN    \undefined \def \showCODEN     #1{\unskip}     \fi
\ifx \showDOI      \undefined \def \showDOI       #1{#1}\fi
\ifx \showISBNx    \undefined \def \showISBNx     #1{\unskip}     \fi
\ifx \showISBNxiii \undefined \def \showISBNxiii  #1{\unskip}     \fi
\ifx \showISSN     \undefined \def \showISSN      #1{\unskip}     \fi
\ifx \showLCCN     \undefined \def \showLCCN      #1{\unskip}     \fi
\ifx \shownote     \undefined \def \shownote      #1{#1}          \fi
\ifx \showarticletitle \undefined \def \showarticletitle #1{#1}   \fi
\ifx \showURL      \undefined \def \showURL       {\relax}        \fi
\providecommand\bibfield[2]{#2}
\providecommand\bibinfo[2]{#2}
\providecommand\natexlab[1]{#1}
\providecommand\showeprint[2][]{arXiv:#2}

\bibitem[\protect\citeauthoryear{??}{ant}{[n.d.]}]%
        {antlr}
 \bibinfo{year}{[n.d.]}\natexlab{}.
\newblock \bibinfo{title}{{ANTLR}}.
\newblock
\newblock
\newblock
\shownote{\url{http://www.antlr.org/} Accessed: July 8, 2025.}


\bibitem[\protect\citeauthoryear{??}{cve}{[n.d.]}]%
        {cve}
 \bibinfo{year}{[n.d.]}\natexlab{}.
\newblock \bibinfo{title}{{Common Vulnerabilities and Exposures}}.
\newblock
\newblock
\newblock
\shownote{\url{https://cve.org/} Accessed: July 8, 2025.}


\bibitem[\protect\citeauthoryear{??}{duc}{[n.d.]}]%
        {duckdb}
 \bibinfo{year}{[n.d.]}\natexlab{}.
\newblock \bibinfo{title}{{DuckDB}}.
\newblock
\newblock
\newblock
\shownote{\url{https://duckdb.org/} Accessed: July 8, 2025.}


\bibitem[\protect\citeauthoryear{??}{ela}{[n.d.]}]%
        {elasticsearch}
 \bibinfo{year}{[n.d.]}\natexlab{}.
\newblock \bibinfo{title}{{Elasticsearch}}.
\newblock
\newblock
\newblock
\shownote{\url{https://www.elastic.co/} Accessed: July 8, 2025.}


\bibitem[\protect\citeauthoryear{??}{neo}{[n.d.]a}]%
        {neo4j}
 \bibinfo{year}{[n.d.]}\natexlab{a}.
\newblock \bibinfo{title}{{Neo4j}}.
\newblock
\newblock
\newblock
\shownote{\url{http://neo4j.com/} Accessed: July 8, 2025.}


\bibitem[\protect\citeauthoryear{??}{neo}{[n.d.]b}]%
        {neo4j-cannot-track-with-filter}
 \bibinfo{year}{[n.d.]}\natexlab{b}.
\newblock \bibinfo{title}{{Neo4j: Conditional Cypher Execution}}.
\newblock
\newblock
\newblock
\shownote{\url{https://neo4j.com/developer/kb/conditional-cypher-execution/} Accessed: July 8, 2025.}


\bibitem[\protect\citeauthoryear{??}{osq}{[n.d.]}]%
        {osquery}
 \bibinfo{year}{[n.d.]}\natexlab{}.
\newblock \bibinfo{title}{osquery}.
\newblock
\newblock
\newblock
\shownote{\url{https://osquery.io/} Accessed: July 8, 2025.}


\bibitem[\protect\citeauthoryear{??}{pos}{[n.d.]}]%
        {postgresql}
 \bibinfo{year}{[n.d.]}\natexlab{}.
\newblock \bibinfo{title}{{PostgreSQL}}.
\newblock
\newblock
\newblock
\shownote{\url{http://www.postgresql.org/} Accessed: July 8, 2025.}


\bibitem[\protect\citeauthoryear{??}{spl}{[n.d.]}]%
        {splunk}
 \bibinfo{year}{[n.d.]}\natexlab{}.
\newblock \bibinfo{title}{{Splunk}}.
\newblock
\newblock
\newblock
\shownote{\url{http://www.splunk.com/} Accessed: July 8, 2025.}


\bibitem[\protect\citeauthoryear{??}{sti}{[n.d.]}]%
        {stix}
 \bibinfo{year}{[n.d.]}\natexlab{}.
\newblock \bibinfo{title}{{Structured Threat Information eXpression}}.
\newblock
\newblock
\newblock
\shownote{\url{http://stixproject.github.io/} Accessed: July 8, 2025.}


\bibitem[\protect\citeauthoryear{??}{sys}{[n.d.]}]%
        {sysdig}
 \bibinfo{year}{[n.d.]}\natexlab{}.
\newblock \bibinfo{title}{{Sysdig}}.
\newblock
\newblock
\newblock
\shownote{\url{http://www.sysdig.com/} Accessed: July 8, 2025.}


\bibitem[\protect\citeauthoryear{??}{aud}{[n.d.]}]%
        {auditd}
 \bibinfo{year}{[n.d.]}\natexlab{}.
\newblock \bibinfo{title}{{The Linux Audit Framework}}.
\newblock
\newblock
\newblock
\shownote{\url{https://github.com/linux-audit/} Accessed: July 8, 2025.}


\bibitem[\protect\citeauthoryear{??}{she}{2014}]%
        {shellshock}
 \bibinfo{year}{2014}\natexlab{}.
\newblock \bibinfo{title}{{CVE}-2014-6271}.
\newblock
\newblock
\newblock
\shownote{\url{https://cve.mitre.org/cgi-bin/cvename.cgi?name=CVE-2014-6271} Accessed: July 8, 2025.}


\bibitem[\protect\citeauthoryear{??}{tar}{2014}]%
        {target}
 \bibinfo{year}{2014}\natexlab{}.
\newblock \bibinfo{title}{{Target Data Breach Incident}}.
\newblock
\newblock
\newblock
\shownote{\url{http://www.nytimes.com/2014/02/27/business/target-reports-on-fourth-quarter-earnings.html} Accessed: July 8, 2025.}


\bibitem[\protect\citeauthoryear{??}{spl}{2016}]%
        {splunkrecursive}
 \bibinfo{year}{2016}\natexlab{}.
\newblock \bibinfo{title}{{Splunk: How to Write a Recursive Search to Build a Tree Structure?}}
\newblock
\newblock
\newblock
\shownote{\url{https://community.splunk.com/t5/Splunk-Search/How-to-write-a-recursive-search-to-build-a-tree-structure/m-p/255819} Accessed: July 8, 2025.}


\bibitem[\protect\citeauthoryear{??}{equ}{2019}]%
        {equifax}
 \bibinfo{year}{2019}\natexlab{}.
\newblock \bibinfo{title}{{Case Equifax Data Breach}}.
\newblock
\newblock
\newblock
\shownote{\url{https://www.ftc.gov/legal-library/browse/cases-proceedings/172-3203-equifax-inc} Accessed: July 8, 2025.}


\bibitem[\protect\citeauthoryear{??}{cyb}{2020}]%
        {cybox}
 \bibinfo{year}{2020}\natexlab{}.
\newblock \bibinfo{title}{{Cyber Observable eXpression}}.
\newblock
\newblock
\newblock
\shownote{\url{https://cyboxproject.github.io/} Accessed: July 8, 2025.}


\bibitem[\protect\citeauthoryear{??}{dar}{2020}]%
        {darpatc}
 \bibinfo{year}{2020}\natexlab{}.
\newblock \bibinfo{title}{{Transparent Computing Engagement 5 Data Release}}.
\newblock
\newblock
\newblock
\shownote{\url{https://github.com/darpa-i2o/Transparent-Computing/blob/master/README.md} Accessed: July 8, 2025.}


\bibitem[\protect\citeauthoryear{??}{cyp}{2021}]%
        {cypher}
 \bibinfo{year}{2021}\natexlab{}.
\newblock \bibinfo{title}{{Cypher Query Language}}.
\newblock
\newblock
\newblock
\shownote{\url{https://neo4j.com/docs/cypher-manual/4.4/introduction/} Accessed: July 8, 2025.}


\bibitem[\protect\citeauthoryear{??}{sql}{2023}]%
        {sql}
 \bibinfo{year}{2023}\natexlab{}.
\newblock \bibinfo{title}{{Information technology — Database languages SQL}}.
\newblock
\newblock
\newblock
\shownote{\url{https://www.iso.org/standard/76583.html} Accessed: July 8, 2025.}


\bibitem[\protect\citeauthoryear{??}{cro}{2024}]%
        {crowdstrike-threat-report}
 \bibinfo{year}{2024}\natexlab{}.
\newblock \bibinfo{title}{Crowdstrike 2024 Global Threat Report}.
\newblock
\newblock
\newblock
\shownote{\url{https://go.crowdstrike.com/rs/281-OBQ-266/images/GlobalThreatReport2024.pdf} Accessed: July 8, 2025.}


\bibitem[\protect\citeauthoryear{??}{ibm}{2024}]%
        {ibm-threat-report}
 \bibinfo{year}{2024}\natexlab{}.
\newblock \bibinfo{title}{IBM X-Force Threat Intelligence Index 2024}.
\newblock
\newblock
\newblock
\shownote{\url{https://newsletter.radensa.ru/wp-content/uploads/2024/03/IBM-XForce-Threat-Intelligence-Index-2024.pdf} Accessed: July 8, 2025.}


\bibitem[\protect\citeauthoryear{??}{pro}{2024}]%
        {procmon}
 \bibinfo{year}{2024}\natexlab{}.
\newblock \bibinfo{title}{{ProcMon}}.
\newblock
\newblock
\newblock
\shownote{\url{https://learn.microsoft.com/en-us/sysinternals/downloads/procmon} Accessed: July 8, 2025.}


\bibitem[\protect\citeauthoryear{??}{pro}{2025}]%
        {provexa-website}
 \bibinfo{year}{2025}\natexlab{}.
\newblock \bibinfo{title}{{Provexa Project Website}}.
\newblock
\newblock
\newblock
\shownote{\url{https://provexa-app.github.io/}.}


\bibitem[\protect\citeauthoryear{??}{bre}{2025}]%
        {breach21}
 \bibinfo{year}{2025}\natexlab{}.
\newblock \bibinfo{title}{{The 20 biggest data breaches of the 21st century}}.
\newblock
\newblock
\newblock
\shownote{\url{https://www.csoonline.com/article/534628/the-biggest-data-breaches-of-the-21st-century.html} Accessed: July 8, 2025.}


\bibitem[\protect\citeauthoryear{Altinisik, Deniz, and Sencar}{Altinisik et~al\mbox{.}}{2023}]%
        {altinisik2023provg}
\bibfield{author}{\bibinfo{person}{Enes Altinisik}, \bibinfo{person}{Fatih Deniz}, {and} \bibinfo{person}{H\"{u}srev~Taha Sencar}.} \bibinfo{year}{2023}\natexlab{}.
\newblock \showarticletitle{ProvG-Searcher: A Graph Representation Learning Approach for Efficient Provenance Graph Search}. In \bibinfo{booktitle}{\emph{ACM SIGSAC Conference on Computer and Communications Security (CCS)}}. \bibinfo{pages}{2247–2261}.
\newblock


\bibitem[\protect\citeauthoryear{Barisic, Amaral, and Goulão}{Barisic et~al\mbox{.}}{2012}]%
        {barisicusabilitydsl}
\bibfield{author}{\bibinfo{person}{Ankica Barisic}, \bibinfo{person}{Vasco Amaral}, {and} \bibinfo{person}{Miguel Goulão}.} \bibinfo{year}{2012}\natexlab{}.
\newblock \showarticletitle{Usability Evaluation of Domain-Specific Languages}. In \bibinfo{booktitle}{\emph{2012 Eighth International Conference on the Quality of Information and Communications Technology}}. \bibinfo{pages}{342--347}.
\newblock
\urldef\tempurl%
\url{https://doi.org/10.1109/QUATIC.2012.63}
\showDOI{\tempurl}


\bibitem[\protect\citeauthoryear{Borders, Springer, and Burnside}{Borders et~al\mbox{.}}{2012}]%
        {chimera}
\bibfield{author}{\bibinfo{person}{Kevin Borders}, \bibinfo{person}{Jonathan Springer}, {and} \bibinfo{person}{Matthew Burnside}.} \bibinfo{year}{2012}\natexlab{}.
\newblock \showarticletitle{Chimera: A Declarative Language for Streaming Network Traffic Analysis}. In \bibinfo{booktitle}{\emph{USENIX Security Symposium (USENIX Security)}}. \bibinfo{pages}{365--379}.
\newblock


\bibitem[\protect\citeauthoryear{Bradshaw, Brazil, and Chodorow}{Bradshaw et~al\mbox{.}}{2019}]%
        {bradshaw2019mongodb}
\bibfield{author}{\bibinfo{person}{Shannon Bradshaw}, \bibinfo{person}{Eoin Brazil}, {and} \bibinfo{person}{Kristina Chodorow}.} \bibinfo{year}{2019}\natexlab{}.
\newblock \bibinfo{booktitle}{\emph{MongoDB: the definitive guide: powerful and scalable data storage}}.
\newblock \bibinfo{publisher}{O'Reilly Media}.
\newblock


\bibitem[\protect\citeauthoryear{Cheng, Bajaber, Tsegai, Song, and Gao}{Cheng et~al\mbox{.}}{2025}]%
        {cheng2025ctinexus}
\bibfield{author}{\bibinfo{person}{Yutong Cheng}, \bibinfo{person}{Osama Bajaber}, \bibinfo{person}{Saimon~Amanuel Tsegai}, \bibinfo{person}{Dawn Song}, {and} \bibinfo{person}{Peng Gao}.} \bibinfo{year}{2025}\natexlab{}.
\newblock \showarticletitle{CTINexus: Automatic Cyber Threat Intelligence Knowledge Graph Construction Using Large Language Models}. In \bibinfo{booktitle}{\emph{IEEE European Symposium on Security and Privacy (EuroS\&P)}}.
\newblock


\bibitem[\protect\citeauthoryear{Cuppens and Ortalo}{Cuppens and Ortalo}{2000}]%
        {lambda}
\bibfield{author}{\bibinfo{person}{Fr{\'e}d{\'e}ric Cuppens} {and} \bibinfo{person}{Rodolphe Ortalo}.} \bibinfo{year}{2000}\natexlab{}.
\newblock \showarticletitle{{LAMBDA}: A Language to Model a Database for Detection of Attacks}. In \bibinfo{booktitle}{\emph{Recent Advances in Intrusion Detection (RAID)}}. \bibinfo{pages}{197--216}.
\newblock


\bibitem[\protect\citeauthoryear{Fang, Gao, Liu, Ayday, Jee, and Wang}{Fang et~al\mbox{.}}{2022}]%
        {fang2022backpropagating}
\bibfield{author}{\bibinfo{person}{Pengcheng Fang}, \bibinfo{person}{Peng Gao}, \bibinfo{person}{Changlin Liu}, \bibinfo{person}{Erman Ayday}, \bibinfo{person}{Kangkook Jee}, {and} \bibinfo{person}{Ting Wang}.} \bibinfo{year}{2022}\natexlab{}.
\newblock \showarticletitle{Back-Propagating System Dependency Impact for Attack Investigation}. In \bibinfo{booktitle}{\emph{USENIX Security Symposium (USENIX Security)}}. \bibinfo{pages}{2461--2478}.
\newblock


\bibitem[\protect\citeauthoryear{Gao, Liu, Choi, Ma, Yang, and Song}{Gao et~al\mbox{.}}{2023}]%
        {gao2023threatkg}
\bibfield{author}{\bibinfo{person}{Peng Gao}, \bibinfo{person}{Xiaoyuan Liu}, \bibinfo{person}{Edward Choi}, \bibinfo{person}{Sibo Ma}, \bibinfo{person}{Xinyu Yang}, {and} \bibinfo{person}{Dawn Song}.} \bibinfo{year}{2023}\natexlab{}.
\newblock \showarticletitle{ThreatKG: An AI-Powered System for Automated Open-Source Cyber Threat Intelligence Gathering and Management}. In \bibinfo{booktitle}{\emph{ACM Workshop on Large AI Systems and Models with Privacy and Safety Analysis (LAMPS)}}. \bibinfo{pages}{1--12}.
\newblock


\bibitem[\protect\citeauthoryear{Gao, Liu, Choi, Soman, Mishra, Farris, and Song}{Gao et~al\mbox{.}}{2021a}]%
        {gao2021security}
\bibfield{author}{\bibinfo{person}{Peng Gao}, \bibinfo{person}{Xiaoyuan Liu}, \bibinfo{person}{Edward Choi}, \bibinfo{person}{Bhavna Soman}, \bibinfo{person}{Chinmaya Mishra}, \bibinfo{person}{Kate Farris}, {and} \bibinfo{person}{Dawn Song}.} \bibinfo{year}{2021}\natexlab{a}.
\newblock \showarticletitle{A System for Automated Threat Intelligence Gathering and Management}. In \bibinfo{booktitle}{\emph{International Conference on Management of Data (SIGMOD)}}. \bibinfo{pages}{2716--2720}.
\newblock


\bibitem[\protect\citeauthoryear{Gao, Shao, Liu, Xiao, Qin, Xu, Mittal, Kulkarni, and Song}{Gao et~al\mbox{.}}{2021b}]%
        {gao2021enabling}
\bibfield{author}{\bibinfo{person}{Peng Gao}, \bibinfo{person}{Fei Shao}, \bibinfo{person}{Xiaoyuan Liu}, \bibinfo{person}{Xusheng Xiao}, \bibinfo{person}{Zheng Qin}, \bibinfo{person}{Fengyuan Xu}, \bibinfo{person}{Prateek Mittal}, \bibinfo{person}{Sanjeev~R Kulkarni}, {and} \bibinfo{person}{Dawn Song}.} \bibinfo{year}{2021}\natexlab{b}.
\newblock \showarticletitle{Enabling Efficient Cyber Threat Hunting with Cyber Threat Intelligence}. In \bibinfo{booktitle}{\emph{International Conference on Data Engineering (ICDE)}}. \bibinfo{pages}{193--204}.
\newblock


\bibitem[\protect\citeauthoryear{Hassan, Guo, Li, Chen, Jee, Li, and Bates}{Hassan et~al\mbox{.}}{2019}]%
        {hassan2019nodoze}
\bibfield{author}{\bibinfo{person}{Wajih~Ul Hassan}, \bibinfo{person}{Shengjian Guo}, \bibinfo{person}{Ding Li}, \bibinfo{person}{Zhengzhang Chen}, \bibinfo{person}{Kangkook Jee}, \bibinfo{person}{Zhichun Li}, {and} \bibinfo{person}{Adam Bates}.} \bibinfo{year}{2019}\natexlab{}.
\newblock \showarticletitle{NODOZE: Combatting Threat Alert Fatigue with Automated Provenance Triage}. In \bibinfo{booktitle}{\emph{Network and Distributed System Security Symposium (NDSS)}}.
\newblock


\bibitem[\protect\citeauthoryear{Inam, Chen, Goyal, Liu, Mink, Michael, Gaur, Bates, and Hassan}{Inam et~al\mbox{.}}{2023}]%
        {inam2023history}
\bibfield{author}{\bibinfo{person}{Muhammad~Adil Inam}, \bibinfo{person}{Yinfang Chen}, \bibinfo{person}{Akul Goyal}, \bibinfo{person}{Jason Liu}, \bibinfo{person}{Jaron Mink}, \bibinfo{person}{Noor Michael}, \bibinfo{person}{Sneha Gaur}, \bibinfo{person}{Adam Bates}, {and} \bibinfo{person}{Wajih~Ul Hassan}.} \bibinfo{year}{2023}\natexlab{}.
\newblock \showarticletitle{{{SoK}}: History is a Vast Early Warning System: Auditing the Provenance of System Intrusions}. In \bibinfo{booktitle}{\emph{IEEE Symposium on Security and Privacy (S\&P)}}. \bibinfo{pages}{2620--2638}.
\newblock


\bibitem[\protect\citeauthoryear{Ji, Lee, Downing, Wang, Fazzini, Kim, Orso, and Lee}{Ji et~al\mbox{.}}{2017}]%
        {ji2017rain}
\bibfield{author}{\bibinfo{person}{Yang Ji}, \bibinfo{person}{Sangho Lee}, \bibinfo{person}{Evan Downing}, \bibinfo{person}{Weiren Wang}, \bibinfo{person}{Mattia Fazzini}, \bibinfo{person}{Taesoo Kim}, \bibinfo{person}{Alessandro Orso}, {and} \bibinfo{person}{Wenke Lee}.} \bibinfo{year}{2017}\natexlab{}.
\newblock \showarticletitle{Rain: Refinable attack investigation with on-demand inter-process information flow tracking}. In \bibinfo{booktitle}{\emph{ACM SIGSAC Conference on Computer and Communications Security (CCS)}}. \bibinfo{pages}{377--390}.
\newblock


\bibitem[\protect\citeauthoryear{Killian, Anderson, Braud, Jhala, and Vahdat}{Killian et~al\mbox{.}}{2007}]%
        {mace}
\bibfield{author}{\bibinfo{person}{Charles~Edwin Killian}, \bibinfo{person}{James~W. Anderson}, \bibinfo{person}{Ryan Braud}, \bibinfo{person}{Ranjit Jhala}, {and} \bibinfo{person}{Amin~M. Vahdat}.} \bibinfo{year}{2007}\natexlab{}.
\newblock \showarticletitle{Mace: Language Support for Building Distributed Systems}. In \bibinfo{booktitle}{\emph{Programming Language Design and Implementation (PLDI)}}. \bibinfo{pages}{179--188}.
\newblock


\bibitem[\protect\citeauthoryear{King and Chen}{King and Chen}{2003}]%
        {backtracking}
\bibfield{author}{\bibinfo{person}{Samuel~T. King} {and} \bibinfo{person}{Peter~M. Chen}.} \bibinfo{year}{2003}\natexlab{}.
\newblock \showarticletitle{Backtracking intrusions}. In \bibinfo{booktitle}{\emph{ACM Symposium on Operating Systems Principles (SOSP)}}. \bibinfo{pages}{223--236}.
\newblock


\bibitem[\protect\citeauthoryear{King, Mao, Lucchetti, and Chen}{King et~al\mbox{.}}{2005}]%
        {backtracking2}
\bibfield{author}{\bibinfo{person}{Samuel~T. King}, \bibinfo{person}{Zhuoqing~Morley Mao}, \bibinfo{person}{Dominic~G. Lucchetti}, {and} \bibinfo{person}{Peter~M. Chen}.} \bibinfo{year}{2005}\natexlab{}.
\newblock \showarticletitle{Enriching Intrusion Alerts Through Multi-Host Causality}. In \bibinfo{booktitle}{\emph{Network and Distributed System Security Symposium (NDSS)}}.
\newblock


\bibitem[\protect\citeauthoryear{Kosar, Mernik, and Carver}{Kosar et~al\mbox{.}}{2012}]%
        {kosardslgpl}
\bibfield{author}{\bibinfo{person}{Tomaž Kosar}, \bibinfo{person}{Marjan Mernik}, {and} \bibinfo{person}{Jeff Carver}.} \bibinfo{year}{2012}\natexlab{}.
\newblock \showarticletitle{Program comprehension of domain-specific and general-purpose languages: Comparison using a family of experiments}.
\newblock \bibinfo{journal}{\emph{Empirical Software Engineering}}  \bibinfo{volume}{17} (\bibinfo{year}{2012}), \bibinfo{pages}{276--304}.
\newblock
\urldef\tempurl%
\url{https://doi.org/10.1007/s10664-011-9172-x}
\showDOI{\tempurl}


\bibitem[\protect\citeauthoryear{Kwon, Wang, Wang, Lee, Lee, Ma, Zhang, Xu, Jha, Ciocarlie, et~al\mbox{.}}{Kwon et~al\mbox{.}}{2018}]%
        {kwon2018mci}
\bibfield{author}{\bibinfo{person}{Yonghwi Kwon}, \bibinfo{person}{Fei Wang}, \bibinfo{person}{Weihang Wang}, \bibinfo{person}{Kyu~Hyung Lee}, \bibinfo{person}{Wen-Chuan Lee}, \bibinfo{person}{Shiqing Ma}, \bibinfo{person}{Xiangyu Zhang}, \bibinfo{person}{Dongyan Xu}, \bibinfo{person}{Somesh Jha}, \bibinfo{person}{Gabriela~F Ciocarlie}, {et~al\mbox{.}}} \bibinfo{year}{2018}\natexlab{}.
\newblock \showarticletitle{{{MCI}}: Modeling-based Causality Inference in Audit Logging for Attack Investigation.}. In \bibinfo{booktitle}{\emph{Network and Distributed System Security Symposium (NDSS)}}.
\newblock


\bibitem[\protect\citeauthoryear{Lee, Zhang, and Xu}{Lee et~al\mbox{.}}{2013}]%
        {lee2013high}
\bibfield{author}{\bibinfo{person}{Kyu~Hyung Lee}, \bibinfo{person}{Xiangyu Zhang}, {and} \bibinfo{person}{Dongyan Xu}.} \bibinfo{year}{2013}\natexlab{}.
\newblock \showarticletitle{High Accuracy Attack Provenance via Binary-based Execution Partition}. In \bibinfo{booktitle}{\emph{Network and Distributed System Security Symposium (NDSS)}}.
\newblock


\bibitem[\protect\citeauthoryear{Liu, Zhang, Li, Jee, Li, Wu, Rhee, and Mittal}{Liu et~al\mbox{.}}{2018}]%
        {liu2018priotracker}
\bibfield{author}{\bibinfo{person}{Yushan Liu}, \bibinfo{person}{Mu Zhang}, \bibinfo{person}{Ding Li}, \bibinfo{person}{Kangkook Jee}, \bibinfo{person}{Zhichun Li}, \bibinfo{person}{Zhenyu Wu}, \bibinfo{person}{Junghwan Rhee}, {and} \bibinfo{person}{Prateek Mittal}.} \bibinfo{year}{2018}\natexlab{}.
\newblock \showarticletitle{Towards a Timely Causality Analysis for Enterprise Security}. In \bibinfo{booktitle}{\emph{Network and Distributed System Security Symposium (NDSS)}}.
\newblock


\bibitem[\protect\citeauthoryear{Loo, Condie, Garofalakis, Gay, Hellerstein, Maniatis, Ramakrishnan, Roscoe, and Stoica}{Loo et~al\mbox{.}}{2006}]%
        {networklang}
\bibfield{author}{\bibinfo{person}{Boon~Thau Loo}, \bibinfo{person}{Tyson Condie}, \bibinfo{person}{Minos Garofalakis}, \bibinfo{person}{David~E. Gay}, \bibinfo{person}{Joseph~M. Hellerstein}, \bibinfo{person}{Petros Maniatis}, \bibinfo{person}{Raghu Ramakrishnan}, \bibinfo{person}{Timothy Roscoe}, {and} \bibinfo{person}{Ion Stoica}.} \bibinfo{year}{2006}\natexlab{}.
\newblock \showarticletitle{Declarative Networking: Language, Execution and Optimization}. In \bibinfo{booktitle}{\emph{ACM SIGMOD International Conference on Management of Data (SIGMOD)}}. \bibinfo{pages}{97--108}.
\newblock


\bibitem[\protect\citeauthoryear{Ma, Zhang, and Xu}{Ma et~al\mbox{.}}{2016}]%
        {ma2016protracer}
\bibfield{author}{\bibinfo{person}{Shiqing Ma}, \bibinfo{person}{Xiangyu Zhang}, {and} \bibinfo{person}{Dongyan Xu}.} \bibinfo{year}{2016}\natexlab{}.
\newblock \showarticletitle{ProTracer: towards practical provenance tracing by alternating between logging and tainting}. In \bibinfo{booktitle}{\emph{Network and Distributed System Security Symposium (NDSS)}}.
\newblock


\bibitem[\protect\citeauthoryear{McMillan}{McMillan}{2013}]%
        {McMillan2013}
\bibfield{author}{\bibinfo{person}{Rob McMillan}.} \bibinfo{year}{2013}\natexlab{}.
\newblock \bibinfo{title}{Definition: Threat Intelligence}.
\newblock \bibinfo{howpublished}{\url{https://www.gartner.com/en/documents/2487216} Accessed: July 8, 2025}.
\newblock


\bibitem[\protect\citeauthoryear{Milajerdi, Eshete, Gjomemo, and Venkatakrishnan}{Milajerdi et~al\mbox{.}}{2019a}]%
        {milajerdi2019poirot}
\bibfield{author}{\bibinfo{person}{Sadegh~M Milajerdi}, \bibinfo{person}{Birhanu Eshete}, \bibinfo{person}{Rigel Gjomemo}, {and} \bibinfo{person}{VN Venkatakrishnan}.} \bibinfo{year}{2019}\natexlab{a}.
\newblock \showarticletitle{Poirot: Aligning attack behavior with kernel audit records for cyber threat hunting}. In \bibinfo{booktitle}{\emph{ACM SIGSAC Conference on Computer and Communications Security (CCS)}}. \bibinfo{pages}{1795–1812}.
\newblock


\bibitem[\protect\citeauthoryear{Milajerdi, Gjomemo, Eshete, Sekar, and Venkatakrishnan}{Milajerdi et~al\mbox{.}}{2019b}]%
        {milajerdi2019holmes}
\bibfield{author}{\bibinfo{person}{Sadegh~M Milajerdi}, \bibinfo{person}{Rigel Gjomemo}, \bibinfo{person}{Birhanu Eshete}, \bibinfo{person}{R Sekar}, {and} \bibinfo{person}{VN Venkatakrishnan}.} \bibinfo{year}{2019}\natexlab{b}.
\newblock \showarticletitle{HOLMES: Real-Time APT Detection through Correlation of Suspicious Information Flows}. In \bibinfo{booktitle}{\emph{IEEE Symposium on Security and Privacy (S\&P)}}. \bibinfo{pages}{1137--1152}.
\newblock


\bibitem[\protect\citeauthoryear{Riddle, Westfall, and Bates}{Riddle et~al\mbox{.}}{2023}]%
        {riddle2023atlasv2atlasattackengagements}
\bibfield{author}{\bibinfo{person}{Andy Riddle}, \bibinfo{person}{Kim Westfall}, {and} \bibinfo{person}{Adam Bates}.} \bibinfo{year}{2023}\natexlab{}.
\newblock \bibinfo{title}{ATLASv2: ATLAS Attack Engagements, Version 2}.
\newblock
\newblock
\showeprint[arxiv]{2401.01341}~[cs.CR]
\urldef\tempurl%
\url{https://arxiv.org/abs/2401.01341}
\showURL{%
\tempurl}


\bibitem[\protect\citeauthoryear{W3C}{W3C}{2008}]%
        {sparql}
\bibfield{author}{\bibinfo{person}{W3C}.} \bibinfo{year}{2008}\natexlab{}.
\newblock \bibinfo{title}{{SPARQL Query Language for RDF}}.
\newblock
\newblock
\newblock
\shownote{\url{https://www.w3.org/TR/rdf-sparql-query/} Accessed: July 8, 2025.}


\bibitem[\protect\citeauthoryear{Wei, Cai, Zhao, Yu, and Meng}{Wei et~al\mbox{.}}{2021}]%
        {wei2021deephunter}
\bibfield{author}{\bibinfo{person}{Renzheng Wei}, \bibinfo{person}{Lijun Cai}, \bibinfo{person}{Lixin Zhao}, \bibinfo{person}{Aimin Yu}, {and} \bibinfo{person}{Dan Meng}.} \bibinfo{year}{2021}\natexlab{}.
\newblock \showarticletitle{Deephunter: A graph neural network based approach for robust cyber threat hunting}. In \bibinfo{booktitle}{\emph{EAI International Conference on Security and Privacy in Communication Networks (SecureComm)}}. \bibinfo{pages}{3--24}.
\newblock


\bibitem[\protect\citeauthoryear{Xu, Wu, Li, Jee, Rhee, Xiao, Xu, Wang, and Jiang}{Xu et~al\mbox{.}}{2016}]%
        {reduction}
\bibfield{author}{\bibinfo{person}{Zhang Xu}, \bibinfo{person}{Zhenyu Wu}, \bibinfo{person}{Zhichun Li}, \bibinfo{person}{Kangkook Jee}, \bibinfo{person}{Junghwan Rhee}, \bibinfo{person}{Xusheng Xiao}, \bibinfo{person}{Fengyuan Xu}, \bibinfo{person}{Haining Wang}, {and} \bibinfo{person}{Guofei Jiang}.} \bibinfo{year}{2016}\natexlab{}.
\newblock \showarticletitle{High Fidelity Data Reduction for Big Data Security Dependency Analyses}. In \bibinfo{booktitle}{\emph{ACM SIGSAC Conference on Computer and Communications Security (CCS)}}. \bibinfo{pages}{504–516}.
\newblock


\end{thebibliography}

\end{document}